\documentstyle{mn}
\input{epsf}
\def\spose#1{\hbox to 0pt{#1\hss}}
\def\simlt{\mathrel{\spose{\lower 3pt\hbox{$\mathchar"218$}}
     \raise 2.0pt\hbox{$\mathchar"13C$}}}
\def\simgt{\mathrel{\spose{\lower 3pt\hbox{$\mathchar"218$}}
     \raise 2.0pt\hbox{$\mathchar"13E$}}}

\def\etal{{\rm et~al. }}
\def\hmpc{\;h^{-1}{\rm Mpc}}
\def\h2mpc{\;h^{-2}{\rm Mpc}}
\def\hmpccc{\;h^{3}{\rm Mpc}^{-3}}

\def\kms{{\rm km\;s}^{-1}}

\def\degree{\nobreak\ifmmode{^\circ}\else{$^\circ$}\fi}

\begin{document}
\title[APM cluster-galaxy correlations]{The APM cluster-galaxy 
cross-correlation function : \\Constraints on $\Omega$ and galaxy bias.}
\author[R.A.C. Croft, G.B. Dalton \& G. Efstathiou]{Rupert A. C. Croft$^{1}$,
Gavin B. Dalton$^2$ \& George Efstathiou$^3$ 
\vspace{1mm}\\
$^1$Department of Astronomy, The Ohio State University,
Columbus, Ohio 43210, USA.\\
$^2$Department of Physics, University of Oxford,  
Keble Road, Oxford, OX1 3RH, UK.\\
$^3$Institute of Astronomy, University of Cambridge,  
Madingley Road, Cambridge, CB3 0HA, UK.\\
}
\maketitle
\begin{abstract}
We estimate   the cluster-galaxy cross-correlation function 
($\xi_{cg}$),
from the APM galaxy and galaxy cluster surveys.
We obtain estimates both in real space from the inversion of projected
statistics and in redshift space using the galaxy and cluster
redshift samples. The amplitude of $\xi_{cg}$ is found to be almost
independent of cluster richness. At large separations, $r \simgt 5 \hmpc$
( $h=H_{0}/100 \kms$, where $H_{0}$ is the 
Hubble constant),
$\xi_{cg}$ has a similar shape to the galaxy-galaxy and cluster-cluster 
autocorrelation functions. 
$\xi_{cg}$ in redshift space can be related to the real space
$\xi_{cg}$ by convolution with an appropriate velocity field model. 
Here we apply a spherical collapse model, which we have tested  against N-body 
simulations, finding that it provides a surprisingly accurate description of
the averaged infall velocity of matter into galaxy clusters. We use this
model to estimate  $\beta$ ($\beta$=$\Omega^{0.6}/b$ where $b$ is the linear 
bias parameter) and find that it tends to overestimate the true result
in simulations by only $\sim10-30\%$.
Application to  the APM results yields  $\beta=0.43$ with $\beta < 0.87$ at
$95\%$ confidence.
This measure is complementary to the estimates made of the density parameter
from larger scale bulk flows and from the virialised regions of clusters on
smaller scales.
We also compare the APM $\xi_{cg}$ and galaxy autocorrelations
directly to the mass correlation and
cluster-mass correlations in COBE normalised simulations
of popular cosmological models and derive two independent
estimates of the galaxy biasing expected as a function of scale. 
This reveals that both low density and critical density cold dark matter
(CDM)  models
require anti-biasing  by a factor $\sim 2$ on scales $r \leq 2\hmpc$
and that the Mixed Dark Matter (MDM)
model is consistent with a constant biasing factor
on all scales.
The critical density CDM model also suffers from the usual deficit of
power on large scales ($r \simgt 20 \hmpc$).
 We use the velocity fields predicted from the different 
models to distort the APM real space cross-correlation function. Comparison
with the APM redshift space $\xi_{cg}$ yields
an estimate of the value of  $\Omega^{0.6}$ needed in each model.
 We find that only the low $\Omega$
model is fully consistent with observations, with MDM marginally
excluded at the $\sim 2
\sigma$ level.

\end{abstract}
\begin{keywords}
Galaxies : Clustering ; Large-scale structure of the Universe ; 
Clusters of galaxies; Cosmology.
\end{keywords}

\section{Introduction}
The spatial 
cluster-galaxy cross-correlation function $\xi_{cg}(r)$ is defined  
so that
the probability dP of finding a galaxy in the volume element dV at a distance
$r$ from a cluster is 
\begin{equation}
\label{dp}
dP=\overline{n}[1+\xi_{cg}(r)]dV,
\end{equation}
where $\overline n$ is the mean space density of galaxies.
 $\xi_{cg}(r)$ is therefore equivalent to the radially averaged
 overdensity profile of galaxies centred on a typical cluster of
 galaxies.  The first measurements of galaxy-cluster correlations were
 made by Seldner \& Peebles (1977) who measured the angular
 cross-correlations of Lick galaxies around Abell clusters. Their
 results suggested that $\xi_{cg}(r)$ was positive and significantly
 different from zero out to large spatial separations of $r \sim
 100\hmpc$.  A more recent analysis by Lilje \& Efstathiou (1988) used
 cluster redshifts to determine the cross-correlation between Abell
 clusters and Lick galaxies as a function of metric separation
 ($w(\sigma)$, see Section 2.2). They found no convincing evidence for
 clustering on scales $r\simgt 20 \hmpc$ and concluded that some of
 the signal seen by Selder and Peebles was due to artificial gradients
 in the Lick catalogue. However, Lilje and Efstathiou did find some
 evidence for more large-scale power than predicted by
 the `standard' ({\it i.e.} $\Omega=1$, $h=0.5$, scale-invariant) CDM
 model. On scales smaller than $r \simlt 10 \hmpc$, the
 real-space $\xi_{cg}(r)$ recovered by inverting the observed form of
 $w(\sigma)$ is well fit by a power law:

\begin{equation}
\label{lickpl}
\xi_{cg}(r)=\left(\frac{r_{0}}{r}\right)^{\gamma},
\end{equation}
with $\gamma=2.2$ and $r_{0}=8.8 \hmpc$.

Direct measurements of the spatial cross-correlation function from
redshift surveys of galaxies and clusters (Dalton, 1992, Efstathiou
1993, Mo, Peacock and Xia ,1993, Moore \etal 1994) have confirmed that
$\xi_{cg}$ has a similar shape to the galaxy-galaxy and
cluster-cluster correlation functions, but with an amplitude roughly
equal to their geometric mean. The analyses of Dalton (1992) and
Efstathiou (1993) were performed using the original APM cluster sample
of Dalton \etal (1992) and the Stromlo-APM galaxies of Loveday \etal
(1992).

 In this paper we use compute $\xi_{cg}$ using the Stromlo-APM galaxy
redshift survey and the sample of 364 clusters of Dalton \etal
(1994b). We investigate its behaviour as a function of cluster
richness and compare it with the predictions of popular comological
models. We also calculate $\xi_{cg}(\sigma,\pi)$, the
cross-correlation as a function of separation along and perpendicular
to the line of sight. The peculiar velocity field around clusters
influences the shape of $\xi_{cg}(\sigma,\pi)$ and is expected to
depend on $\Omega$, and the mass to light ratio around clusters, so
that we can extract some information about the density parameter and
galaxy biasing from our measurements.  We also carry out some specific
comparisons with N-body simulations (so including non-linear effects)
to investigate how the biasing of galaxies is expected to vary as a
function of scale.

\section{Cluster-galaxy correlations in the APM survey}

\subsection{The APM data samples}
In this paper we will use three different data samples, the
APM angular galaxy catalogue, the APM-Stromlo galaxy redshift survey,
and the APM cluster redshift survey. The APM galaxy survey (Maddox
\etal 1990a, Maddox \etal 1990b, Maddox, Efstathiou and Sutherland
1996) consists of angular positions and other information, but not
redshifts, for over 2 million galaxies with a b$_{J}$ magnitude limit
of 20.5.  The Stromlo-APM redshift survey (Loveday 1990, Loveday \etal
1992a, Loveday \etal 1992b) is a survey of $1787$ galaxies
randomly sampled at a rate of 1 in 20 from all APM galaxies with
b$_{J}$ brighter  than 17.15. To construct the APM cluster sample, an
automated procedure was used to select clusters from the angular APM
survey (Dalton 1992, Dalton \etal 1997).  Cluster redshifts were then
measured, and used to construct an original redshift catalogue of 190
clusters (Dalton \etal 1992) and an extension to 364 clusters (Dalton
\etal 1994). It is the latter cluster catalogue which we will use in
this paper. In our analyses we will use angular clustering and
its inversion to obtain real space 3 dimensional information, using
the parent APM galaxy survey and the cluster redshift survey (Section
2.2). We will also measure the 3 dimensional clustering directly from
the APM-Stromlo redshift survey and the cluster redshift survey
(Section 2.3).

\subsection{The projected cross-correlation function}
The simplest statistic which can be used to constrain $\xi_{cg}(r)$ is
the angular cross-correlation function, $w_{cg}(\theta)$. However, the
inversion of this quantity  to find  $\xi_{cg}(r)$ tends to be rather
 unstable. Since we are using the APM cluster
redshift survey as our cluster sample we can make use of the cluster redshift
information to determine the projected cross-correlation function,
$w_{cg}(\sigma)$, as defined by Lilje \& Efstathiou (1988), where  $\sigma
=cz \theta/H_0$ is the metric separation of a cluster with redshift
$z$ and a galaxy at angular distance $\theta$ from the cluster 
centre.
As our estimator for $w_{cg}(\sigma)$ we use the standard estimator, 
\begin{equation}
\label{estimator}
w_{cg}(\sigma) = \frac{N_{ran}}{N_{gal}}\frac{N_{CG}(\sigma)}{N_{CR}(\sigma)} 
-1,
\end{equation}
where CG and CR denote cluster--galaxy and cluster--random pairs,
respectively. We account for the window function of the APM Galaxy
Survey on a plate-by-plate basis by generating a random catalogue for
a single plate with 100000 random points and excising the regions
masked from the survey as the catalogue is used for each plate in
turn. With this method, we can use individual galaxy positions from the
survey data rather than binned cell counts and so we can measure
$w_{cg}(\sigma)$ accurately at small scales. On larger
scales we have checked that this method does not introduce large-scale
power into our determination of $w_{cg}(\sigma)$ by comparing with
the results obtained using cell counts for galaxies and a single
random catalogue for the whole survey (Dalton 1992). A similar method has
been used more recently in the analysis of the Durham--UKST galaxy
redshift survey (Ratcliffe \etal, 1997).

\begin{figure}
\centering
\vspace{8.0cm}
\includegraphics{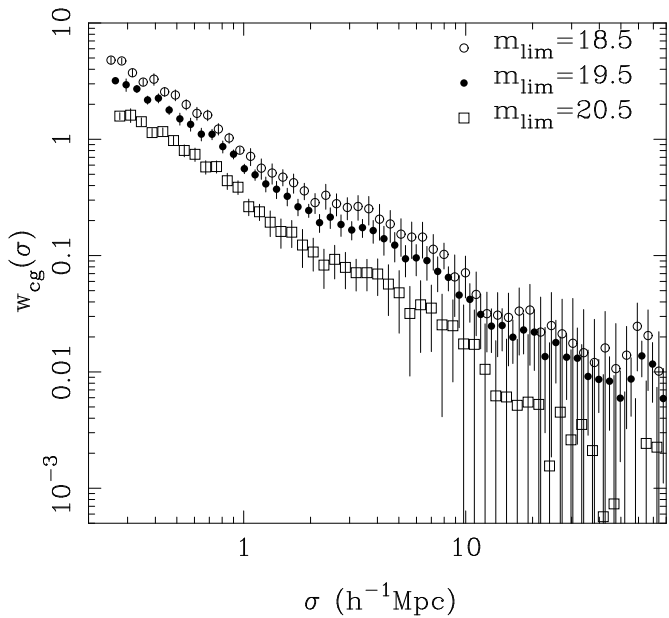}
\caption[xicgrich]{The projected cluster-galaxy cross-correlation function
for APM galaxies and APM clusters. Results are shown using 
APM galaxies with different magnitude limits as indicated 
in the figure. The error bars are determined 
from the scatter in the results derived from  four nearly equal area
zones of the APM survey.

\label{sigcg}}
\end{figure}

The data for $w_{cg}(\sigma)$ are shown in Figure~1 for the cluster
sample cross-correlated with all galaxies to three different magnitude
limits. The error bars shown are obtained by dividing the APM survey
region into four quadrants and determining the error on the mean from
the scatter between the four zones. Given the estimate of the depth of
the Lick catalogue obtained by Maddox \etal (1990a), we would expect
the points for $m_{lim}=18.5$ to correspond to the results for
$w_{cg}(\sigma)$ obtained by Lilje \& Efstathiou (1988). A comparison
of Figure~1 to Figure~9c of that paper reveals good agreement over the
range of $\sigma$ for which $w_{cg}$ can be measured reliably.

\begin{figure}

\centering
\vspace{8.0cm}
\includegraphics{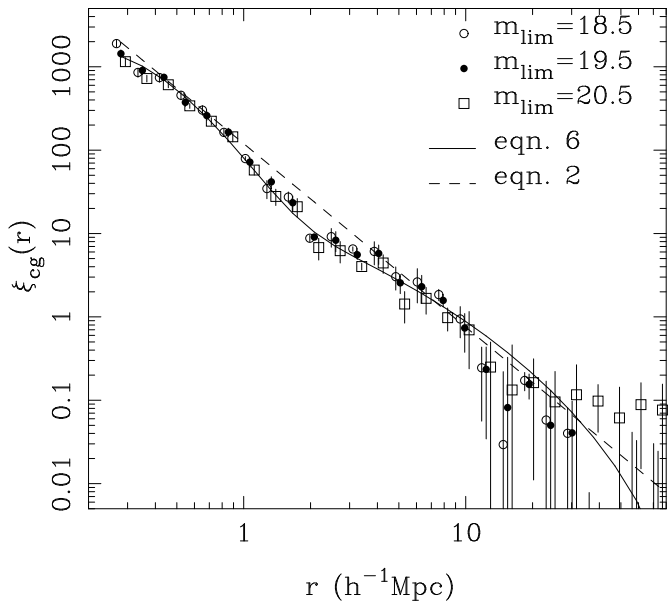}
\caption[xicgrich]{The real space cluster-galaxy cross-correlation function
for APM galaxies and APM clusters. Results are shown for different magnitude
limits, and the error bars have been calculated as described in the caption
for Figure 1. For clarity, the m$_{lim}=18.5$ and m$_{lim}=20.5$ points 
have been  slightly offset in the $r$ direction. We also show a 
fit to the data with form given by Equation \ref{fit1} and the
power law (Equation \ref{lickpl}) which Lilje \& Efstathiou find
is a good fit to the cross-correlation of Abell clusters and Lick counts.  
 \label{xicgreal}} 
\end{figure}

The inversion of projected clustering information in the form of
$w_{cg}(\sigma)$ to the three-dimensional statistic 
 $\xi_{cg}(r)$ involves a weighted summation
of the $w_{cg}(\sigma)$ points (Saunders, Rowan-Robinson  \& Lawrence, 1992):
\begin{eqnarray}
&\xi_{cg}(r)=\frac{-1}{\pi B}\sum_{j\geq i}\frac{w(\sigma_{j+1})-w(\sigma_{j})}
{\sigma_{j+1}-\sigma_{j}} 
 \ln{\left(\frac{\sigma_{j+1}+\sqrt{\sigma_{j+1}^{2}-
\sigma_{i}^{2}}}{\sigma_{j}+
\sqrt{\sigma_{j}^{2}-\sigma_{i}^{2}}}\right)}&,
\end{eqnarray}
where $r=\sigma_{i}$. The factor $B$ in Equation 4  accounts for the
 difference in the selection functions of the clusters and galaxies 
and is defined as follows (Lilje \& Efstathiou 1988):
\begin{equation}
B=\frac{\sum_{i} \psi(y_{i})}{\sum_{i}(1/y_{i})\int_{0}^{\infty}
\psi(x) x^{2} dx}.
\end{equation}
Here $\psi$ is the selection function of the galaxy survey and
$y_{i}$ is the redshift of cluster $i$.
The selection function $\psi$  was evaluated using the
the luminosity function parameters
obtained from the Stromlo-APM survey by Loveday \etal (1992).
 Again we show error bars
obtained  by inverting the $w_{cg}(\sigma)$ estimates from four
quadrants of the survey for each magnitude range. The data show
excellent agreement between the three different magnitude limited
galaxy samples used, but are not
well represented by a single power law.

For the velocity field analysis in Section 4, we will use a
fit to the real space APM cross-correlation function.
 We choose to fit an  arbitrary function  which
is the sum of an exponentially truncated power law and
the linear theory correlation function shape of a scale-invariant
CDM model with $\Gamma=\Omega h =0.2$ 
 (denoted as $\xi_{\Gamma=0.2}(r)$ below) normalised so that
$\sigma_8=1$\footnote{Where $\sigma_8$ denotes the {\it rms}
amplitude of the mass fluctuations in spheres of radius $8 \hmpc$.}:

\begin{equation}
\label{fit1}
\xi_{cg}(r)=\left(\frac{r_{c}}{r}\right)^{\gamma}e^{\beta r}e^{\eta/r}+
\alpha~\xi_{\Gamma=0.2}(r).
\end{equation}

The parameter combination $r_{c}=11.7$, $\gamma=2.6$, $\eta=0.6$, $\beta=1.7$
and $\alpha=1.8$ gives a reasonable fit to the data for all magnitude 
bins and is plotted as a solid line on Figure 2.  
A fit is necessary because the  noise level in the real space
cross-correlation function becomes rather large for $r\simgt10 \hmpc$.
The shape of the fit on these large scales is motivated by the shape of
the redshift space $\xi_{cg}$ (Section 2.2). We also plot on the
same figure the power-law fit  (given by Equation~\ref{lickpl})
which Lilje and Efstathiou (1988)
find is a good approximation to the  real space
cross-correlation function derived from Abell clusters and Lick counts.
We can see that on large scales, there is no evidence for any significant
excess of power over this fit. The APM results therefore
 support the conclusions
of Lilje and Efstathiou (1988) summarized  in Section 1.
Our APM results are in agreement with the cross-correlation
of APM clusters and Edinburgh-Durham Sky Survey galaxies carried
out by Merch\'{a}n \etal (1997), although their estimated errors are large.
It is  useful to note that on small scales, $r \simlt 2 \hmpc$, our
error bars on $\xi_{cg}(r)$ are very small, so that we will be able to
draw some interesting conclusions about galaxy biasing on small
scales from a comparison with theoretical models (Section 5).

\subsection{The redshift space cross-correlation function}

We estimate $\xi_{cg}(s)$, where $s$ represents
the separation of cluster-galaxy pairs in redshift space,
using the APM sample 
of 364 clusters of  Dalton \etal (1994b) and the $\sim 2000$ galaxies
with redshifts from the APM-Stromlo bright galaxy redshift survey 
(Loveday \etal 1992a). 
The calculation of  $\xi_{cg}(s)$ differs from the evaluation
of the cluster-cluster correlation function (Dalton \etal 1994b)
as the galaxy selection function falls very steeply with increasing
distance. Weights must be therefore be applied to the galaxy-cluster pairs
to recover the mimimum variance estimate of $\xi_{cg}(s)$ .
This optimal weighting (at least on scales for which $\xi_{cg}(s) \leq 1$) 
can be shown to be (see Efstathiou 1988, Loveday 1990) 
\begin{equation} 
\label{wt}
w_{ij}\simeq1/[1+4\pi n(r_{i}) J^{cg}_{3}(s_{ij})],
\end{equation}
where
\begin{equation}
J^{cg}_{3}(s_{ij})=\int_{0}^{s_{ij}}\xi_{cg}(x)x^{2}dx,
\end{equation}
$r_{i}$ is the distance from the observer to galaxy $i$ and $s_{ij}$ the 
separation of galaxy $i$ and cluster $j$. To use the formula we must predict
 roughly what $J^{cg}_{3}(s)$ will be - here we use the 
weighting function resulting from a linear theory CDM power spectrum with
$\Gamma=0.2$ and an amplitude twice that of $\xi_{gg}(s)$ measured for
APM-Stromlo galaxies (Loveday \etal 1992). After cross-correlating 
the two catalogues to find all
galaxy-cluster pairs, we cross-correlate the clusters with  a catalogue of
100000 random points. This random catalogue has the same  
boundaries and selection function as the galaxy sample. We then use the 
standard estimator to find $\xi_{cg}(s)$:  

\begin{equation}
\xi_{cg}(s)=\frac{N_{ran}}{N_{gal}} \frac{N_{CG}(s)}{N_{CR}(s)} -1,
\end{equation}
where $N_{CG}(s)$ and $N_{CR}(s)$
 are the galaxy-cluster and cluster-random pairs in the
bin interval centred on $s$ and each are weighted using $w_{ij}$ from
Equation~\ref{wt}.

Results for the 364 ${\cal R}=50$ clusters are shown as triangles in
 Fig. 3.
 It can be seen that the
there is some curvature in the plot, with a definite break at 
$s \sim 30-50 \hmpc $. The curve crosses $\xi_{cg}(s)=1$ at
roughly  $9 \hmpc$, which is intermediate between the behaviour of the galaxy
auto-correlation function,
$\xi_{gg}(s)$ ($s_{0} \simeq 5 \hmpc$, in the notation
of Equation~\ref{lickpl},  Loveday \etal 1992)
 and the cluster auto-correlation function $\xi_{cc}(s)$
($s_{0} \simeq 14 \hmpc$, Dalton \etal 1994).
 The error bars, calculated from Poisson 
statisics and the number of cluster-galaxy pairs in each bin, are relatively
small, indicating that  $\xi_{cg}(s)$ will be an interesting statistic to
compare with theoretical models.
We can compare the results for ${\cal R}=50$ clusters with the solid line 
in this figure, which is the fit to the real space cross-correlation
function (Equation 6). 
For the moment, we will note that  $\xi_{cg}(s)$ is marginally higher than
 $\xi_{cg}(r)$ on large scales, boosted by streaming motions
 (Kaiser 1987) 
and smaller for separations less than $\sim 4 \hmpc$ due to the effect of
the cluster velocity dispersion. Uncertainties in the cluster redshifts 
probably play a part in depressing the amplitude of $\xi_{cg}(s)$ on
small scales, as for many cluster
we  have redshifts for only 2 or 3 galaxies (each with their own measurement
errors). The error in the cluster centre of mass velocity could, therefore,
be as  much as a few hundred $\kms$.  Dalton $\etal$ (1994a) have compared  
results for APM clusters with many ($>10$) measured galaxy redshifts
to the redshifts of the brightest galaxy in each cluster. 
The rms scatter between the two
values is $512~ \kms$, which should be higher than the error on our
cluster redshifts as we use $\geq 2$ galaxy
redshifts per cluster.  

Moore \etal (1994) have found that $\xi_{cg}$ for IRAS galaxies and
Abell clusters is insensitive to cluster richness. We have repeated
this type of analysis using our APM samples and the results are
plotted in Figure 3.  We have increased the lower richness cutoff from
${\cal R}=50$ (the full sample) up to ${\cal R}=80$, but as the
results show there is no detectable change in the shape or amplitude
of $\xi_{cg}(s)$, given the errors. Our different samples have space
densities $\overline{n_{c}}= 3.5\times 10^{-5}\hmpccc$ (${\cal
R}=50$), $2.2\times 10^{-5} \hmpccc$ (${\cal R}=60$), $9.3\times
10^{-6} \hmpccc$ (${\cal R}=70$) and $4.7\times 10^{-6}\hmpccc$
(${\cal R}=80$). We might expect the cross-correlation
function of richer clusters to have a higher amplitude, at least on small 
scales, as cluster richness should be related
to $\xi_{cg}$ within the cluster selection radius.  This was discussed
by Seldner and Peebles (1977), who found that the effect was smaller
than expected for Abell clusters. This is probably because the distance
indicator used by Abell depends on cluster richness; for
clusters at a given apparent distance, the richer objects would
actually be further away, thus depressing the amplitude
of the  angular cross-correlation
function.  These problems should not affect the APM
cluster sample as it has been designed so that cluster richness does
not affect apparent distance (Dalton $\etal$ 1997).
However, the situation here is complicated by the fact that we are
measuring redshift space clustering. Rich clusters will have their clustering
signal smeared out due to their high velocity dispersions, so that the
amplitude of clustering will be more heavily depressed on small scales
than for poorer clusters. The underlying situation on small scales is
 therefore not entirely obvious. On larger scales 
($r \simgt 1 \hmpc$) though, our results show 
that the amplitude of $\xi_cg(s)$ really does depend only very weakly
on cluster richness. We will show in the next section that this is consistent
with model predictions.

\begin{figure}

\centering
\vspace{7.0cm}
\includegraphics{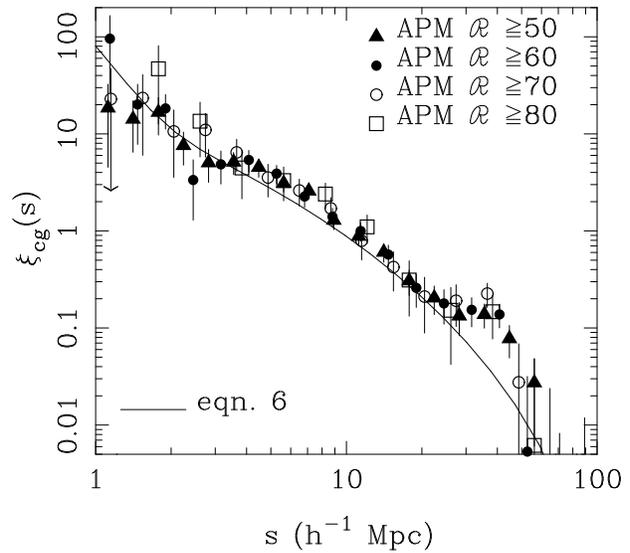}
\caption[xicgrich]{The spatial cluster-galaxy cross-correlation function
for APM-Stromlo galaxies and samples of APM clusters with different
lower richness limits. All cluster subsamples are drawn from the sample
of 364 clusters (APM ${\cal R} \geq 50$), so that 243, 113 and 60 clusters
are used in the calculation of the curves for ${\cal R} \geq 60$,
${\cal R} \geq 70$ and ${\cal R} \geq 80$ clusters respectively.
The  subsamples have mean intercluster separations of
$36 \hmpc$ (${\cal R} \geq 60$), $48 \hmpc$ ( ${\cal R} \geq 70$)
and $59 \hmpc$ ( ${\cal R} \geq 80$).
The solid line is a fit (with form given by Equation 6) to the real
 space $\xi_{cg}$ results for the APM
${\cal R} \geq 50$ clusters.
 \label{xicgrich}}
\end{figure}

We have calculated $\xi_{cg}$ for the APM sample in redshift space as
a function of pair separation along the line of sight ($\pi$) and
perpendicular to the line of sight ($\sigma$). The effects of peculiar
velocities, which distort the pair separations in the $\pi$ direction
are evident in the results for our sample plotted in Figure 4. We can
see elongation present on small scales which is caused by random
galaxy velocities and redshift measurement errors. On larger scales,
we can see a break in the contours around $\sigma\simeq 6 \hmpc,
\pi\simeq 10 \hmpc$ which could be caused by the infall region around
the cluster. This coherent infall should cause compression of the
contours along the $\sigma$ axis on larger scales (Kaiser 1987, Lilje
\& Efstathiou 1989), but there does not seem to be obvious evidence
of this in Figure 4.

The velocity field around clusters should be dependent on the mass
distribution, and therefore on the value of
$\Omega$.  If we have a simple model for how the two main effects
present in velocity field arise (small scale dispersion and large
scale infall) we should be able to use distortions in
$\xi_{cg}(\sigma,\pi)$ to derive information on $\Omega$ and the
amplitude of mass fluctuations. To do this, we need to know $\xi_{cg}$
in real space and to have a model which describes the behaviour of
galaxy velocities around clusters. We apply both of these to our
$\xi_{cg} (\sigma,\pi)$ data from the APM survey in Section 4.3
below. We will first examine the predictions of theoretical models
using N-body simulations and use them to develop a model of the velocity 
field around clusters.

\begin{figure}
\centering
\vspace{7.5cm}
\includegraphics{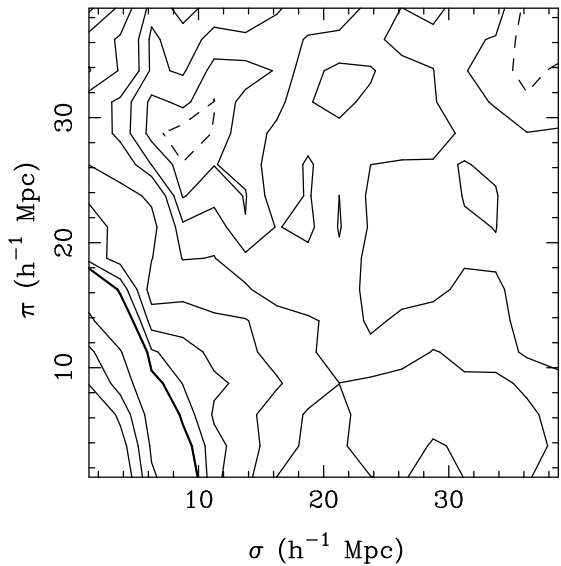}
\caption[xicgsp]{The APM cluster-galaxy cross-correlation function 
$\xi_{cg}(\sigma, \pi)$ (calculated using the full sample of 364 clusters)
shown as a function of pair separations perpendicular to the line
of sight ($\sigma$) and
along the line of sight ($\pi$). Contour levels are at $\xi_{cg}= 
4, 3, 2, 1, 0.8, 0.6, 0.4, 0.2, 0.1, 0., -0.05$. The contour
level at  $\xi_{cg}=1$ is shown by the heavy line; 
negative contours are plotted as
dashed lines. $\xi_{cg}(\sigma, \pi)$ was calculated using 16 bins in 
$\sigma$ and  $\pi$ in the
interval $0-40 \hmpc$. For clarity, before plotting the results in this
figure they were smoothed using a moving window average ($3 \times 3$ bins).
 \label{xicgsp}}
\end{figure}

\section{$\xi_{cg}$ from simulations of cosmological models}

\subsection{Clusters and galaxies in simulations}

Our predictions of the form of the cluster-galaxy correlation function
 in cosmological models come from studying N-body simulations.  We use
 catalogues of clusters constructed from simulations which have been
 used in previous papers to study the cluster-cluster correlation
 function (Croft \& Efstathiou 1994, Dalton \etal 1994b).  The
 simulations are of three different spatially flat universes, two CDM
 models and one Mixed Dark Matter (MDM) model.  One set of simulations
 is of ``standard'' CDM (SCDM), with $\Omega_{0}=1, h=0.5$ and the
 other is of low density CDM (LCDM) with $\Omega_{0}=0.2, h=1$ and a
 cosmological constant $\Lambda$,  where $\Lambda=0.8 \times
 3H_{0}^{2}$.  The power spectra for the SCDM and LCDM models are as
 given Efstathiou, Bond \& White (1992). For the MDM model, we used
 the form given in Klypin et al. (1993) with $\Omega_0=1$, $h=0.5$,
 and a massive neutrino component contributing $\Omega_\nu = 0.3$. We
 assume scale invariant primordial fluctuations for all models.

Each simulation
contains $10^{6}$ particles in a box of comoving side-length $30000 \kms$
and was run using a P$^{3}$M $N$-body code 
(Efstathiou \etal 1985). We use 5 realisations 
of each model with different random phases.

\begin{figure*}
\centering
\vspace{8.0cm}
\includegraphics{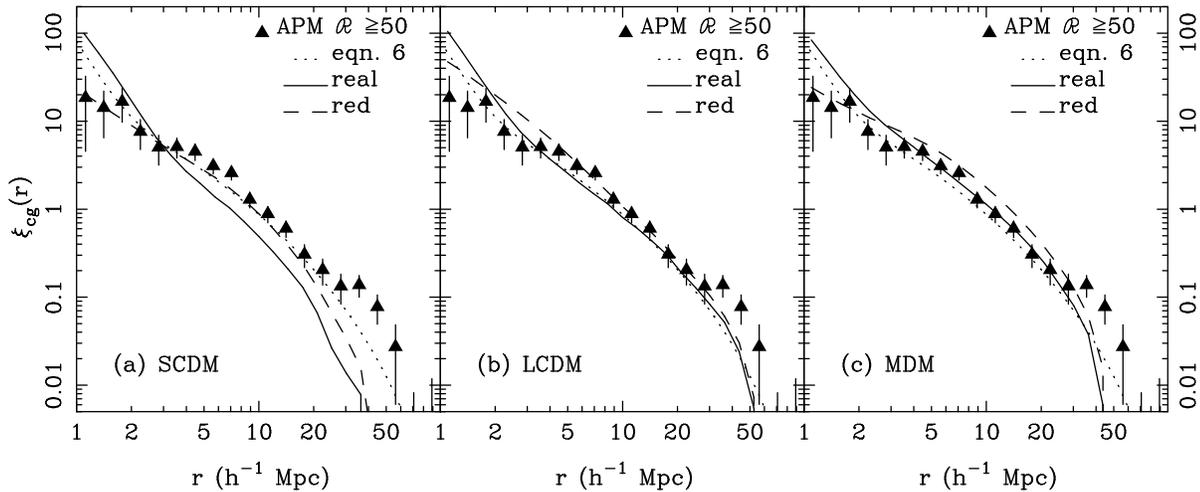}
\caption[xicgrs]{The cluster-mass cross-correlation function for
simulations of (a) Standard CDM, (b) Low Density CDM and (c) Mixed
Dark Matter (see text for the parameters of these models).  Results are plotted in real space (solid lines) and in
redshift space (dashed lines).  All models are normalised to be
consistent with the amplitude of fluctuations measured by COBE
(ignoring tensor modes), so
that $\sigma_{8}= 1.0 $ for panels (a) and (b) and $\sigma_{8}= 0.67 $
in panel (c). The model curves in panel (c) have been scaled upwards
by a biasing factor of 1.5 in order to make the model consistent with
the amplitude of galaxy fluctuations. Also plotted in each panel (as
triangles) is the cross-correlation function of APM Stromlo galaxies
and the APM cluster sample described in Section 2.3. The error bars on these
points have been calculated using Poisson statistics. The dotted lines
show a fit (Equation 6) to the real space cluster-galaxy correlation function,
consistent with measurements from the APM survey and also with the
cross-correlation of Abell clusters and Lick counts.
\label{xicgrs}}
\end{figure*}

In this Section we use simulated cluster catalogues constructed from
the N-body simulations to have the same mean separation as the APM
sample ($d_{c}=30 \hmpc$). The clusters are identified from
simulations using a friends-of-friends algorithm to select candididate
centres. The mass enclosed within a fixed radius (in this case $0.5
\hmpc$) of the centre of mass is computed and the clusters are ordered
by mass. Finally, a mass limit is applied to generate a cluster
catalogue of a specified mean space density.  The procedure is
described in more detail in Croft \& Efstathiou (1994).  We note here
that as long as the clusters are defined to be collapsed objects, the
set of objects identified in the simulations is insensitive to the
selection criterion. For example, Gazta\~{n}aga, Croft \& Dalton
(1995) identify clusters as high peaks in the density field smoothed
with small filters and recover essentially the same catalogue of
clusters (for a variety of filter sizes) as our percolation algorithm.
We can also reasonably expect the positions and mass rankings of
galaxy clusters in our simulations to be insensitive to the details of
the galaxy formation process, which in the real Universe would turn a
large agglomeration of dark matter into a galaxy cluster. These
properties of galaxy clusters make them especially useful for
elucidating the nature of galaxy biasing and galaxy peculiar
velocities. In essence they consist of a set of fixed reference points
around which the galaxy overdensity profiles and galaxy velocity
profiles can be compared directly with observations.

We do not attempt to carry out any sort of direct identification of
galaxies in the simulations.  Instead, we calculate the
cluster-particle corrrelation function, $\xi_{c\rho}$.  The APM
$\xi_{cg}$ observations can be compared to $\xi_{c\rho}$ to determine
what sort of galaxy biasing is needed in each model.  According to
linear perturbation theory, the amplitude of $\sigma_8$ grows in
proportion to the growth rate $D(t)$ of linear density perturbations
(see Peebles 1980, Section 10). The amplitude of the two-point
correlation function of the mass fluctuations thus grows as
$\sigma_{8}^{2} \propto D^2(t)$ on scales on which linear theory is
applicable.  On the other hand, the rich cluster two-point correlation
function, is almost independent of time (see Croft \& Efstathiou 1994)
because clusters are rare objects and are strongly biased compared to
the mass distribution. We therefore expect the linear theory growth
rate of $\xi_{c\rho}$ to lie between these two cases, and to be
proportional to $\sigma_{8}$, as in the high-peak model of Bardeen
\etal 1986.  This means that the choice of output time (and hence
normalisation of the model) will make some difference to our
results. The normalisation we have used is consistent with the
amplitude of fluctuations inferred from COBE microwave background
temperature anisotropies (Wright \etal 1994) so that $\sigma_{8}=1.0$
for SCDM and LCDM and $\sigma_{8}=0.67$ for MDM. If we assume that
$\sigma_{8}$ for APM galaxies is close to 1.0 (Gazta\~{n}aga 1995) and
density evolution on this scale has been linear, then $\xi_{c\rho}$
for SCDM and LCDM should be very nearly equal to $\xi_{cg}$.  In the
case of MDM, when plotting our results we merely scale the curve
upwards by a constant linear biasing factor, $1.0/\sigma_{8}$.  A
discussion of more complicated biasing, including variations with
spatial scale which might be required in some models, is deferred to
Section 5.

\subsection{Results}

\begin{figure*}
\centering
\vspace{8.0cm}
\includegraphics{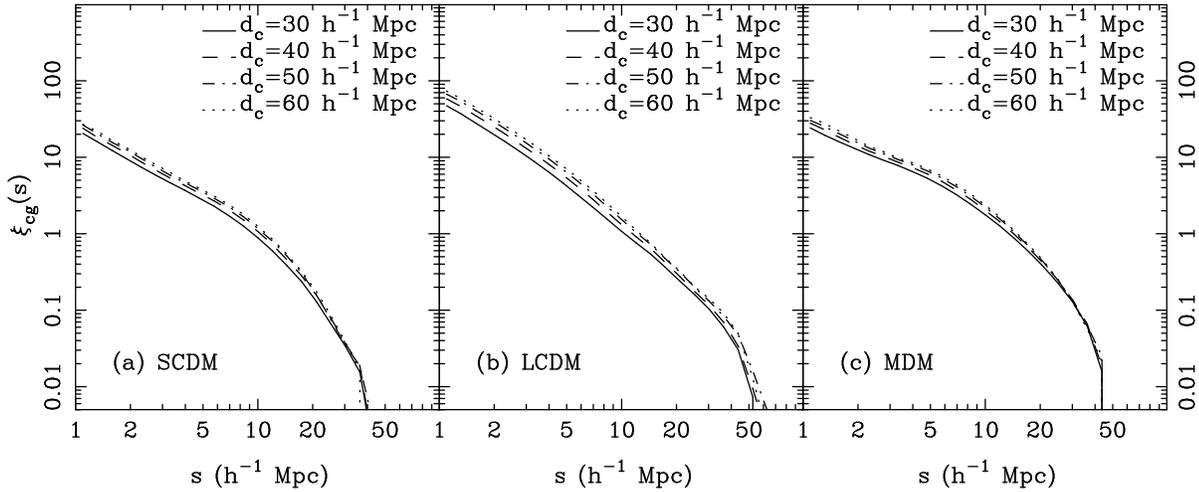}
\caption[xicgrs]{The cluster-mass cross-correlation function in redshift
space for clusters with different lower mass limits  in 
simulations of (a) Standard CDM, (b) Low density CDM and (c) Mixed Dark Matter.
The curves are labelled with the mean intercluster separation of each sample.
 \label{xicgrs1}}
\end{figure*}

Our results for these three models are shown in Figure 5, both in real
space (solid lines) and redshift space (dashed lines). Concentrating
on the real space results first, it can be seen that $\xi_{cg}(r)$
exhibits a sudden change of slope on small scales in all cases. The
SCDM and LCDM models have the steepest slopes for this part of the
curve, which we might expect, given that they have the most small
scale power.  The redshift space results show the usual depression on
small scales and amplification on large scales, both effects being
largest in the case of the two $\Omega=1$ models, which have the
largest particle velocities. As far as a comparison with the APM
$\xi_{cg}(r)$ is concerned, we can see that the usual discrepancy on
large scales with SCDM is evident, but that the shape on these scales
is consistent with LCDM. Indeed for $r>3 \hmpc$ LCDM appears to give a
good fit in both real and redshift space.  On smaller scales, MDM
appears to give a reasonable fit to the shape of both the real space
results (Equation 6) and those in redshift space. We should be cautious
about drawing firm conclusions from this, as the MDM model has been
simulated without including the thermal velocities of massive
neutrinos and this may affect the density profiles and internal
structure of the clusters. The linearly biased MDM model does also
seem to have a rather high amplitude on intermediate scales,
particularly in redshift space (a point that we will return to in
Sections 4 and 5). In any case, it is possible that the efficiency of
galaxy formation is different near clusters leading to scale dependent
biasing. We will investigate this possibility in Section 5.

The richness dependence of $\xi_{cg}$ in the models is shown in  
Figure 6, where we plot  $\xi_{cg}(s)$ (in redshift
space) for simulated clusters
with a similar range of space densities to the 4 different APM samples
shown in Figure 3 . There is a very weak richness dependence in the amplitude
of of $\xi_{cg}$, which becomes even weaker on large 
scales ($s\geq 25\hmpc$). This weak dependence of $\xi_{cg}$ with cluster
richness is compatible with the observational results for the APM samples
presented in Figure 3, which show no significant dependence of $\xi_{cg}$
with cluster richness.

\section{Redshift space distortions and $\xi_{cg}(\sigma,\pi)$}

\subsection{The spherical infall model}

There exists a regime between streaming of galaxies
on large-scales and the virialised 
region of clusters which is important in modelling observations of 
$\xi_{cg}(\sigma,\pi)$.   
As the density enhancement within a few $\hmpc$ of clusters is greater
than unity, linear theory is not expected to describe the velocity field 
accurately.  However, if we assume that clusters are spherically 
symmetric, a solution 
for the non-linear collapse of the system can be found which is exact
before orbit crossing takes place. The solution is obtained by treating 
a proto-cluster with a top-hat profile as if it were an isolated 
Friedmann universe with its own value of $\Omega_{0}$ (see e.g. Reg\"{o}s \&
Geller 1989
for details). Here, we use a good  approximation to the
exact solution due to Yahil (1985) (also used by Lilje \& Efstathiou 1989),
who gives the following expression:
\begin{equation}
v^{non-lin}_{infall}(r)=-\frac{1}{3}\Omega_{0}^{0.6}
H_{0}r\frac{\delta(r)}{(1+\delta(r))^{0.25}},
\end{equation}
where $\delta(r)$ is the overdensity inside radius $r$, 
\begin{equation}
\delta(r)= \frac{3 J^{c\rho}_{3}(r)}{r^{3}}, \qquad J^{c\rho}_3(r)
= \int_0^r \xi_{c\rho}(x) x^2\; dx.
\end{equation}
 By way of comparison, the linear theory prediction of the infall velocity
is
\begin{equation}
v^{lin}_{infall}(r)=-\frac{1}{3}\Omega_{0}^{0.6} H_{0}r\delta(r).
\end{equation} 
Galaxy surveys provide with information on the overdensity
of galaxies inside radius $r$ and not of the mass. Some assumptions 
about the relationship between  galaxies and mass are therefore required
to infer a value of  $\Omega_{0}$ from a measurement of $v_{infall}(r)$.
Here we use the simple linear biasing picture, so that $J^{c\rho}_{3}(r)=
 J^{cg}_{3}(r)/b$. We therefore parametrise our infall model using
$\beta=\Omega_{0}^{0.6}/b$. As even the non-linear velocity model is
not expected to hold in the cores of clusters, near and in the virialised
region, we choose to truncate the expression for $v^{non-lin}_{infall}(r)$
with an exponential $e^{-\delta(r)/\delta_{cut}}$, with $\delta_{cut}=50$.
(See the comparisons with N-body simulations in the next subsection
for a justification of this value of $\delta_{cut}$.)

One further ingredient in our velocity model is the random velocity
dispersion about the smooth infalling flow.  To make things as simple
as possible, we assume a velocity dispersion independent of distance
from the galaxy to the cluster, and independent of direction (whether
transverse to the line between galaxy and cluster, or parallel to it,
for example). We also assume that the velocities are drawn from a
Gaussian distribution. The one-dimensional velocity dispersion will
therefore be parametrised by one number, $\sigma_{v}$.

\subsection{Direct tests of spherical infall
 on the velocity field in simulations.}

\begin{figure*}
\centering
\vspace{7.5cm}
\includegraphics{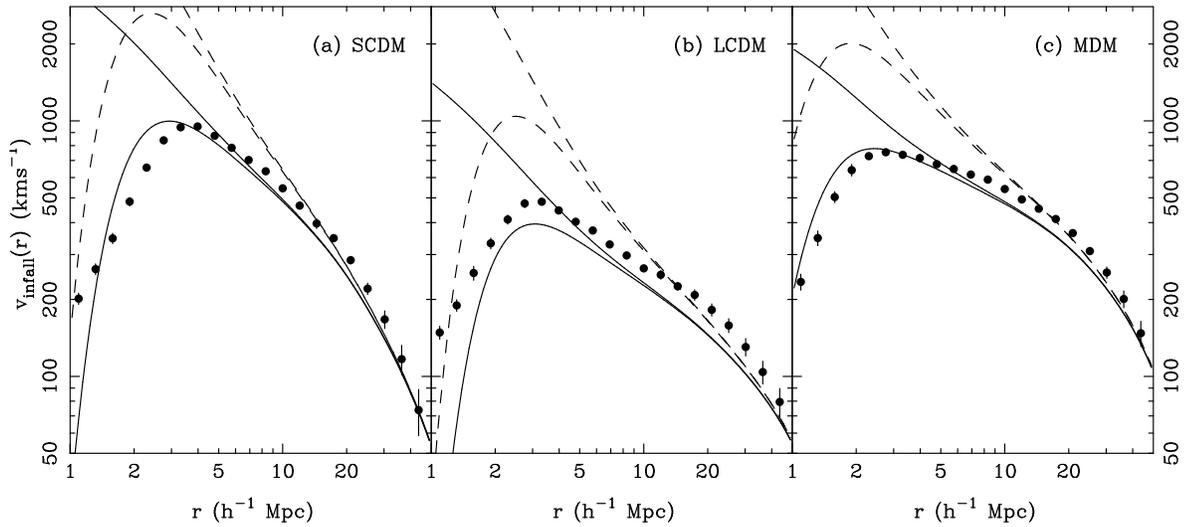}
\caption[vi]{The mean infall velocity of particles around simulated
clusters as a function of radius. The filled circles in each panel
correspond to the infall velocities found in N-body simulations of (a)
SCDM, (b) LCDM and (c) MDM. The error bars show the error on the mean
calculated from the scatter in an ensemble of 5 realisations. In each
panel we have also plotted the predictions of linear theory (dashed
lines) and a non-linear spherical infall model (solid lines), as
described in the text.  The two pairs of dashed and solid lines show
the effect of including and not including an exponential truncation at
high overdensities as described in Section 4.1.  \label{vi1}}
\end{figure*}

 In our analysis we need to assume that the velocity field predicted
from the average cluster density profile is equivalent to the average
cluster velocity field. This is a non-trivial assumption and must be
tested in some way (in our case we will use simulations) before we can
make any claims for the reliablity of our results.  The spherical
infall model has previously been used in many studies of the infall
regions around individual rich clusters, particularily Virgo
(e.g. Yahil, Sandage \& Tammann 1980, Yahil 1985). However, the
assumption of spherical symmetry for any individual cluster is
difficult to justify empirically. In contrast, the average cluster
profile measured by $\xi_{cg}(r)$ possesses spherical symmetry by
construction.  Here we test the validity of our dynamical
approximations by comparing our predicted velocity fields 
directly with the velocity fields measured around N-body clusters.
We present results both for the mean infall velocity
as a function of radius and the dispersion about the mean infall.

\begin{figure*}
\centering
\vspace{7.2cm}
\includegraphics{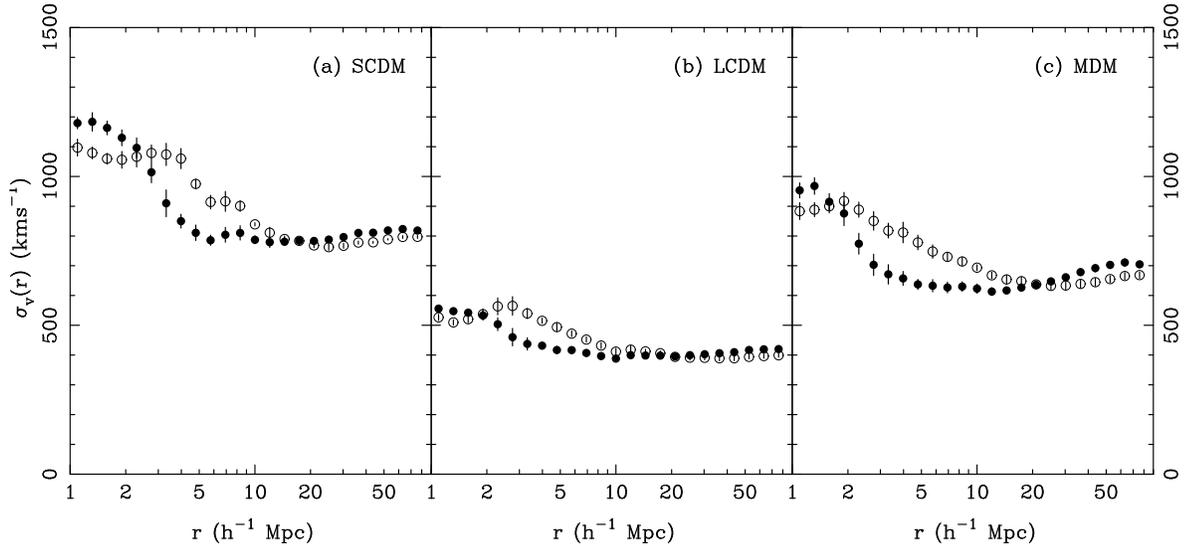}
\caption[vi]{The one-dimensional velocity dispersion of particles
around simulated clusters as a function of radius. The filled circles
in each panel correspond to the dispersion in the component of
 relative velocity along the line from particle to cluster  
(ie. dispersion about the mean
streaming motion plotted in Figure 7).  The open circles show the
one-dimensional dispersion in the transverse direction. We plot the
same models as in previous diagrams: (a) SCDM, (b) LCDM and (c)
MDM. The error bars again show the error on the mean calculated from
the scatter in an ensemble of 5 realisations. 
\label{vi}}
\end{figure*}

The average infall velocities of particles, are plotted as a function
 of distance from the cluster centre in Figure 7 (filled circles).  We
 also plot the predictions of spherical infall (solid lines) and
 linear theory (dashed lines). The rapid decrease of infall velocity
 at small $r$ in the simulations is due to particles reaching the
 boundary of the virialised region.  We have roughly approximated this
 effect in our velocity models by the exponential truncation described
 in Section 4.1. The pairs of dashed and solid curves show
 the velocity models with and without the exponential truncation.
 Because of the somewhat arbitrary nature of this
 truncation, we will restrict our quantitative use of the velocity
 model to the region $r > 2.5 \hmpc$.  Evidently, 
 the spherical non-linear model provides a much better match to
 the N-body results than the linear. Over much of
 the range of density contrasts relevant to  our study 
 of the infall region, the  non-linear velocity model 
 underpredicts the infall 
 by roughly constant factor, but this is only $\sim 30\%$ for
 LCDM and $\sim 10-20\%$ for SCDM and MDM.  At larger radii, $r
 \geq 20 \hmpc$ we  see that both linear and non-linear models
 begin to match the N-body results even more closely.  We therefore
 expect that when we come to estimate $\beta$ from distortions in the
 cross-correlation function our results will be fairly close to the
 true value.

The underprediction of the spherical infall model is in agreement with
results found by Van Haarlem (1992) and Diaferio and Geller (1996).
As the velocity field in the immediate vicinity of galaxy clusters has
the potential to be complicated, there have been many mechanisms
suggested that might disrupt simple spherical infall. Among these are
the presence of shear which should speed up collapse (Hoffman 1986,
Lilje \& Lahav 1991), which does not appear to be happening here, and
formation of substructure and/or ellipticity of the proto-cluster
which slows infall (e.g. van Haarlem \& van de Weygaert 1993).  The
study of clusters in the SCDM scenario carried out by Villumsen \&
Davis (1986) differs from ours in that they analysed the velocity
field around individual clusters. Here we are dealing with the
averaged velocity around all clusters. 

\begin{figure*}
\centering
\vspace{12.5cm}
\includegraphics{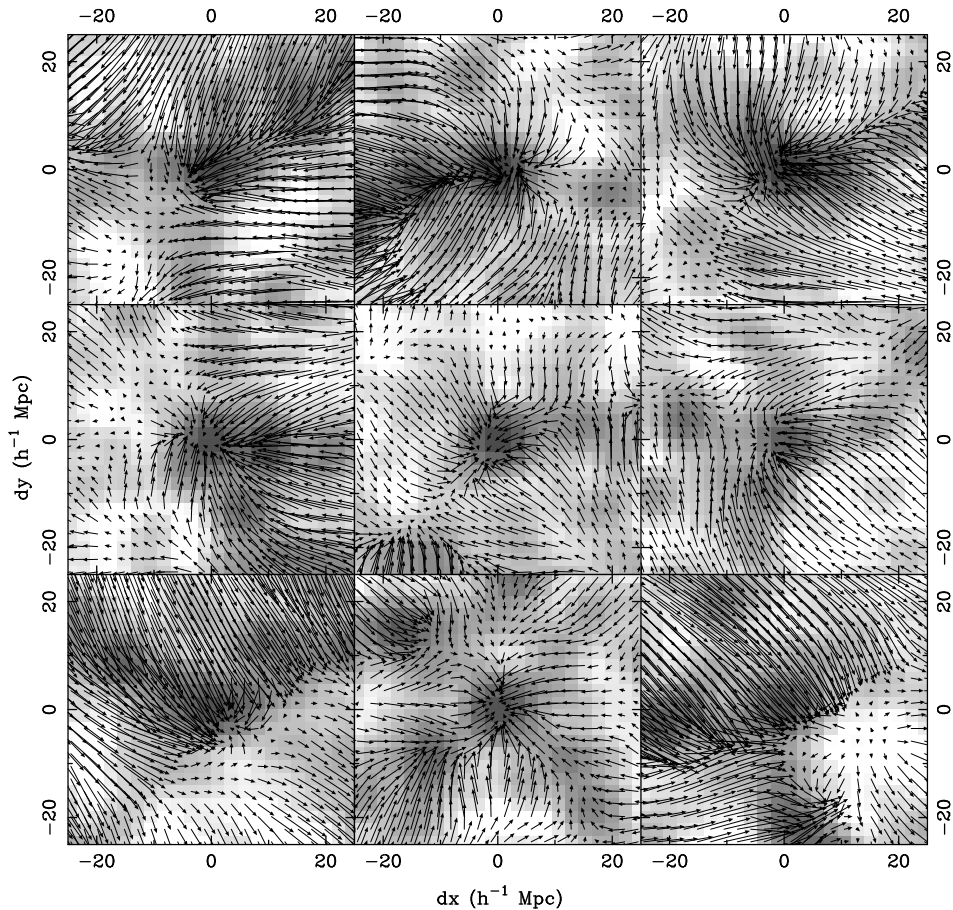}
\caption[md]{The distribution of particles (with their projected velocities)
surrounding 9 clusters taken from a simulation of a standard CDM universe (
with $\sigma_{8}$ =1.0). 
The clusters plotted correspond to 
those ranked by mass as numbers 2,4,8,16,32,64,128,256 and 512 from  a
box of size $300 \hmpc$. In each panel, the cluster in question is located in
the centre, and the velocity field of particles (in the 
rest frame of the cluster) around the cluster has been 
assigned to a grid using a $2 \hmpc$ Gaussian kernel. The X and Y 
components of this velocity field in a slice through the centre
of the cluster are shown as arrows. The particles have also been assigned to a
grid using the same  kernel and a slice through the resulting density field
is also  shown on each panel as a grayscale (linear in density).
 \label{md1}}
\end{figure*}

Figure 8 is a plot of the 1-dimensional relative velocity 
dispersion as a function of distance from particle to cluster in our three
cosmological models. The filled symbols show the dispersion about the mean
infall motion along the line from particle to cluster
 and the open circles show
the one-dimensional dispersion in the component of relative velocity 
transverse to this line.
It can be seen 
that our assumption that $\sigma_{v}$ is independent of distance and
direction turns out to be surprisingly good 
 The plots also show that the transverse
$\sigma_{v}$ is slightly higher than the radial value, in the infall region
between $r\sim 2 \hmpc$ and $r\sim 10 \hmpc$. This might be the signature
of some sort of ``previrialisation'' occuring (Davis \& Peebles 1977,
Peebles 1993).

\begin{figure*}
\centering
\vspace{12.5cm}
\includegraphics{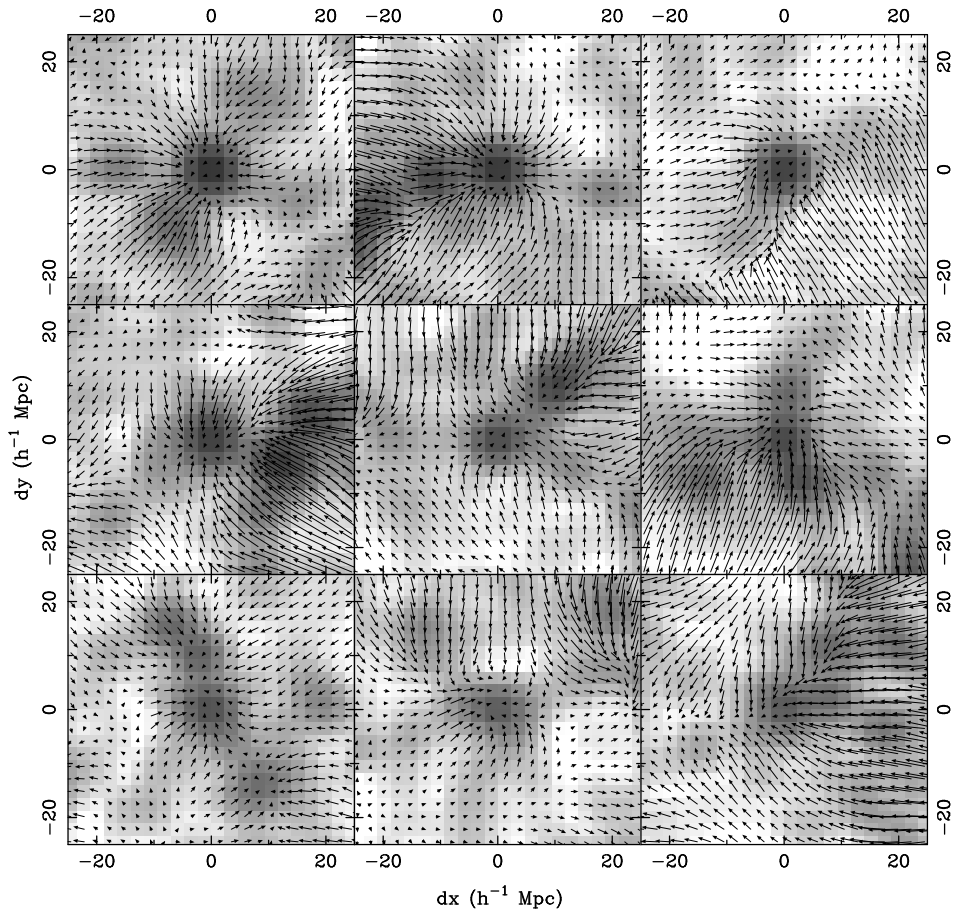}
\caption[md]{As for Fig 9 but for an LCDM simulation (with $\sigma_{8}=1.0$).
 \label{md}}
\end{figure*}

\begin{figure*}
\centering
\vspace{12.5cm}
\includegraphics{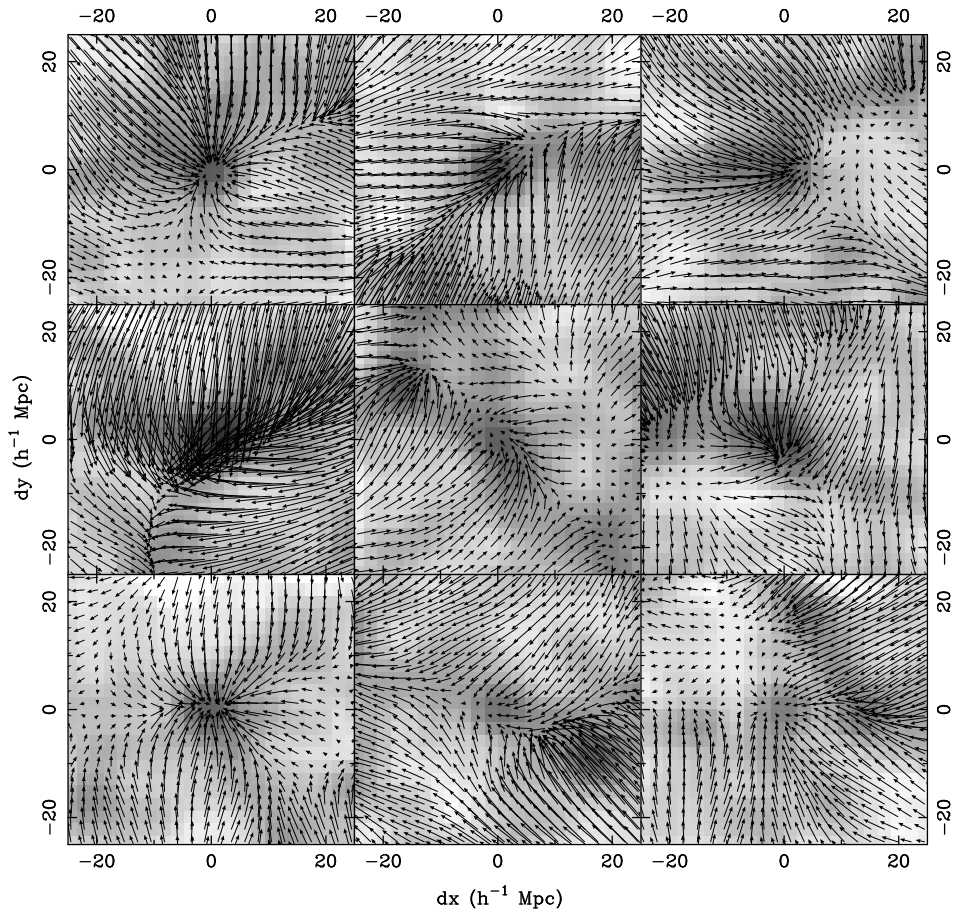}
\caption[md2]{As for Fig 9 but for a MDM simulation 
(with $\sigma_{8}=0.67$).
 \label{md}}
\end{figure*}

To illustrate the complexity of the velocity field around the
 simulated clusters, and to see if we can understand why some
 scenarios fit the spherical infall picture slightly better than
 others, we have plotted the particle velocities around a sample of
 individual clusters. In Figures 9 to 11 we show the x and y
 components of the smoothed peculiar velocity field (a 2$\hmpc$
 Gaussian filter was used) around 9 clusters for each model, in the
 rest frame of the cluster (plotted at the centre of each panel).
 We have also plotted the density field smoothed with the same filter
 as a grayscale.  We can reach several conclusions from studying these plots:

\noindent
$\bullet$ The magnitude of the velocity field around MDM clusters 
(and to a lesser extent SCDM) is anisotropic - there are several
clusters with small arrows on one side and large ones on the other.\\ 
$\bullet$ The larger coherence in the velocity field of the MDM model 
compared to the two CDM models is evident.
Also obvious is the greater magnitude of the velocities in this model.\\
$\bullet$There is substucture in the density field for all models
and the cluster we are interested in at the centre of the panel is
often part of an elongated system.
The MDM model also has a noticeably smoother density field on small
scales.\\
$\bullet$ All models contain some clusters where the main flow pattern
is not centered on the cluster but continues past it,
and others have more complex and non-radial flows. 
The pictures do not inspire confidence in the idea of simple spherically 
symmetric infall around each cluster. It seems that this picture is
only likely to apply after some sort of averaging is carried out, 
at least around each cluster (e.g. as in Villumsen \& Davis 1986)
or around a statistically defined sample of clusters,  as described 
in this paper. It is also worth considering that these models 
 do have a relatively
high amplitude of mass fluctuations, being normalized to COBE rather
 than to give
the correct abundance of galaxy clusters (White, Efstathiou \&
Frenk 1993). It is possible that lower amplitude models would conform better
to a smooth infall picture.

Obvious signatures of radial infall, such as overdensity ``caustics'' at the
radius of turnaround from the Hubble flow (Reg\"{o}s \& Geller 1989) have
been difficult to find in observations of real clusters and in N-body
simulations (van Haarlem \etal 1993).  This does not seem surprising
given the complexity evident in these plots. This is 
consistent with  the work of Colberg \etal (1997) who have shown that
matter infalling onto individual clusters in simulations does so from
specific directions which are correlated with the surrounding large
scale structure.  It is possible that modelling the velocity and
density fields as triaxial ellipsoids (van Haarlem \& van de Veygaert
1993) may help in the understanding of the flow. In any case the
accuracy of our spherically averaged approximation is good enough to
permit a quantitative analysis of the redshift space distortions.

\subsection{Estimates of $\beta$ and $\sigma_{v}$}

We are now in the position to make some predictions of how the real
space $\xi_{cg}$ will be distorted as a function of $\beta$ and $\sigma_{v}$.
To do this, we shall convolve $\xi_{cg}(r)$ with our velocity model,
so that $\xi_{cg}$ in redshift space, $\xi_{cg}(\sigma,\pi)$, is given by
(Peebles 1993, Section 20)
\begin{eqnarray}
&\xi_{cg}(\sigma,\pi)=(\sqrt{2\pi}\sigma_{v})^{-1}\int^{\infty}_{-\infty}
[\exp{(v-[(v^{non-lin}_{infall}(r)y)/\sigma_{v}])^{2}} & \nonumber \\
&\qquad \times(1+\xi_{cg}(r)) -\exp{(-v^{2}/\sigma_{v})^{2}}]\;dv,&
\label{convolve}
\end{eqnarray}
where $r=\sqrt{(\sigma^{2}+\pi^{2})}$ and $y=\sqrt{(r^{2}-\sigma^{2})}$.
To model the observed estimates of $\xi_{cg} (\sigma, \pi)$, we model
$\xi_{cg}(r)$ in Equation 13 with the fitting function of Equation
6  which is consistent with the real
space cross-correlation function of the APM sample (Section 2.2). We
also test Equation 13 using the N-body simulations and empirical
estimates of $\xi_{cg}(r)$ calculated in real space (plotted in
Fig. 5).

\begin{figure*}
\centering
\vspace{9.7cm}
\includegraphics{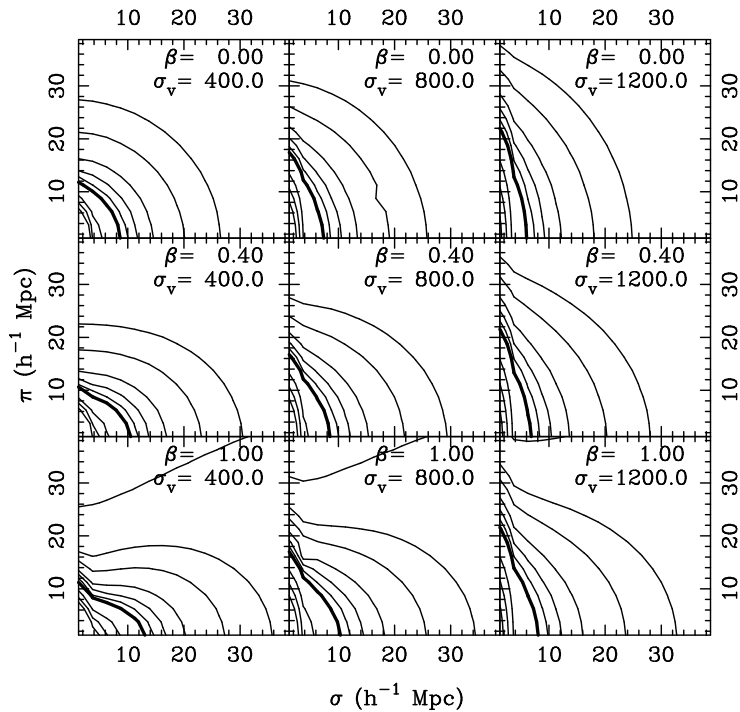}
\caption[modsp]{A smooth fit (Equation 6)
to the  observed cluster-galaxy cross-correlation function
$\xi_{cg}(r)$  convolved with a velocity model (spherical infall
and constant velocity dispersion- see text) to produce
a predicted $\xi_{cg}(\sigma, \pi)$  plot. We show results for 3 different
values of $\beta$ and 3 different values of the velocity dispersion,
$\sigma_{v}$. 
Contour levels are the same as in Figure 4
 \label{modsp}}
\end{figure*}

To place constraints on $\beta$ and $\sigma_{v}$ we calculate the 
expected   $\xi_{cg} (\sigma,\pi) $ for a closely spaced grid of values,
over the range $\beta=0-1.5$ and $\sigma_{v}=0-1500 \kms$. 
$\xi_{cg} (\sigma,\pi) $ is calculated at $16\times16$ points,
spanning the range of values for $\pi$ and $\sigma$ plotted in Figure 4.
This is done by convolving the real space $\xi_{cg}(r)$
 with our velocity model using Equation 13.
In Figure 12 we show contour plots for a few sample values of $\beta$
and $\sigma_{v}$.
We then find which of our grid of models has the smallest
$\chi^{2}$  by comparing with the observed $\xi_{cg} (\sigma,\pi) $.
We estimate confidence bounds on $\beta$ and $\sigma_{v}$ from the
distribution of $\Delta \chi^{2}$ over the [$\beta, \sigma_{v}$] plane. 
As the infall model does not give an accurate prediction for the
velocity in the virialised  region, we have chosen to exclude from the
fit  the  $\xi_{cg} (\sigma,\pi)$ point with the smallest value
of $\sigma$ and $\pi$. Thus  we exclude any  cluster-galaxy pairs 
with  both $\sigma$ and $\pi < 2.5 \hmpc$.

\begin{figure*}
\centering
\vspace{11.0cm}
\includegraphics{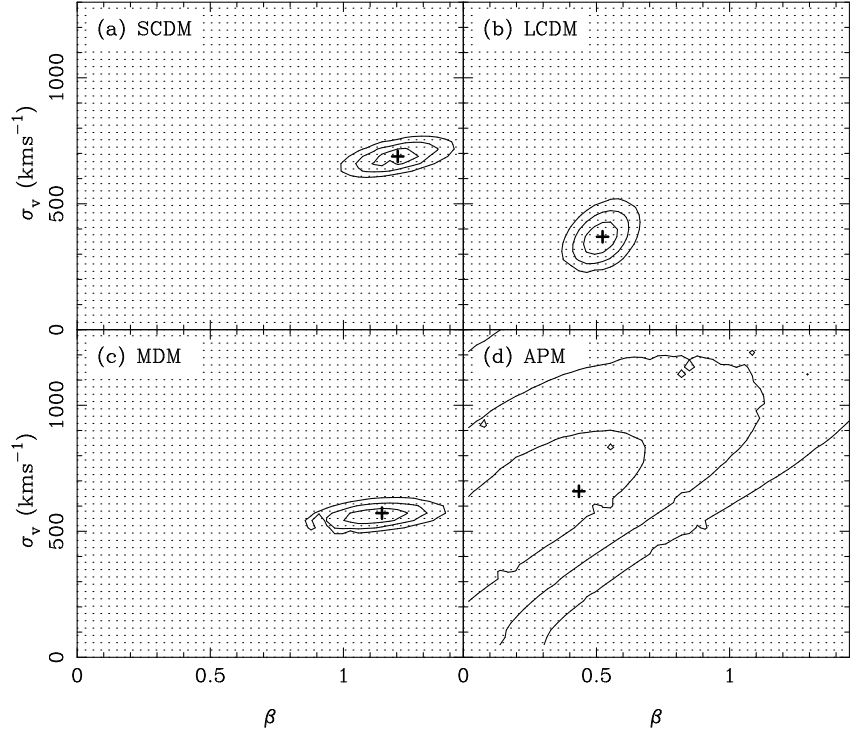}
\caption[xichi2]{
Contours of constant $\Delta \chi^{2}$ corresponding to $68\%, 95\%$ and
$99.7\%$ confidence limits on values of $\sigma$ (the 1 dimensional velocity
dispersion) and $\beta$ ($\simeq \Omega^{0.6}/b$) found by fitting a velocity
model to $\xi_{cg}(\sigma, \pi)$. In panels (a) to (c) we show the results 
found from clusters identified in 5 ensembles each of simulations of
the SCDM, LCDM and MDM models respectively. In panel (d) we show results 
for the APM survey (using the $\xi_{cg}(\sigma, \pi)$ plotted in Figure 4).
In each panel the values of $\sigma$ and $\beta$ for which the $\chi^{2}$ was
evaluated are shown by dots, and the best fit values are indicated by a cross.
 \label{xichi2}}
\end{figure*}

The $\chi^{2}$ contours are shown 
in Figure 13, where we have tested the  SCDM, LCDM and
MDM simulations. For the test on the N-body models,
we have used all ($\approx 1000$) clusters in each box
(and all $10^{6}$ particles), as well as averaging over
5 realisations. We  can  see straight away that we are 
predicting  values, for $\beta$ at least,
that are consistent with the results of our direct comparison of
infall velocities plotted in Figure 7.
 For the LCDM model,
for which the true value is $\beta=0.39$ (Lahav \etal 1991),
 we find  $\beta=0.52^{+0.06}_{-0.09}$(95\%
confidence limits).  For SCDM and MDM we measure   $\beta=1.20^{+0.09}
_{-0.12}$ and  $\beta=1.14^{+0.12}_{-0.18}$,  repectively.
The error bars on $\beta$ given here are computed by 
 marginalisation over all values
of $\sigma_{v}$. The confidence limits on the pair of parameters  
$\sigma_v$ and $\beta$ are plotted as  contours in Figure 13. 
This diagram shows that given enough clusters we should
be able to distinguish between high and low $\beta$ models. This constraint
on $\beta$ from the infall region around clusters (most of the signal comes from
 $r < 10 \hmpc$) is complementary both to measurements of the central
mass of clusters from the virial theorem and to larger scale bulk flow
observations (e.g. Loveday \etal 1996).
 As the infall method tends to systematically overestimate
 $\beta$ by a small amount, any measurement which gives a particularly low
 value, for example one inconsistent with high $\Omega$ models,
will be especially interesting.

The  values of $\sigma_{v}$ we measure using the maximum
likelihood fits  are $370^{+60}_{-90} \kms$, $690^{+60}_{-60} \kms$,
and  $570^{+30}_{-60} \kms$ for LCDM, SCDM and MDM respectively.
These are again $95 \%$ confidence limits obtained
after marginalising over all values of
$\beta$.  Comparison with Fig. 8 reveals that these values are in good
agreement with the velocity dispersion of particles in the outer parts
of clusters. 

We now turn to the determination of  $\beta$ and $\sigma_{v}$ from the APM
sample. The results of the maximum likelihood fitting are shown in panel
 (d) of Figure 13. We can see immediately that we have much larger statistical
errors than in our simulations, as there 
are only 364 clusters and $\sim$ 2000 galaxies in the APM sample.
The value of $\beta$ we obtain is $0.43$, with $\beta < 0.87$ at 95\%
confidence.
 The tail of probablilties towards rather high values of $\sigma_{v}$ and the
 best fit $\sigma_{v}$ inferred,
$660 \kms$ ($\sigma_{v} < 1070 \kms$ at 95\%
confidence)  are probably a consequence of uncertainties in the cluster
redshifts.  There is not therefore a good constraint on the
velocity dispersion in the outer regions of clusters from the APM sample.

\section{Model-dependent comparisons: biasing as a function of scale.}

\begin{figure*}
\centering
\vspace{8.0cm}
\includegraphics{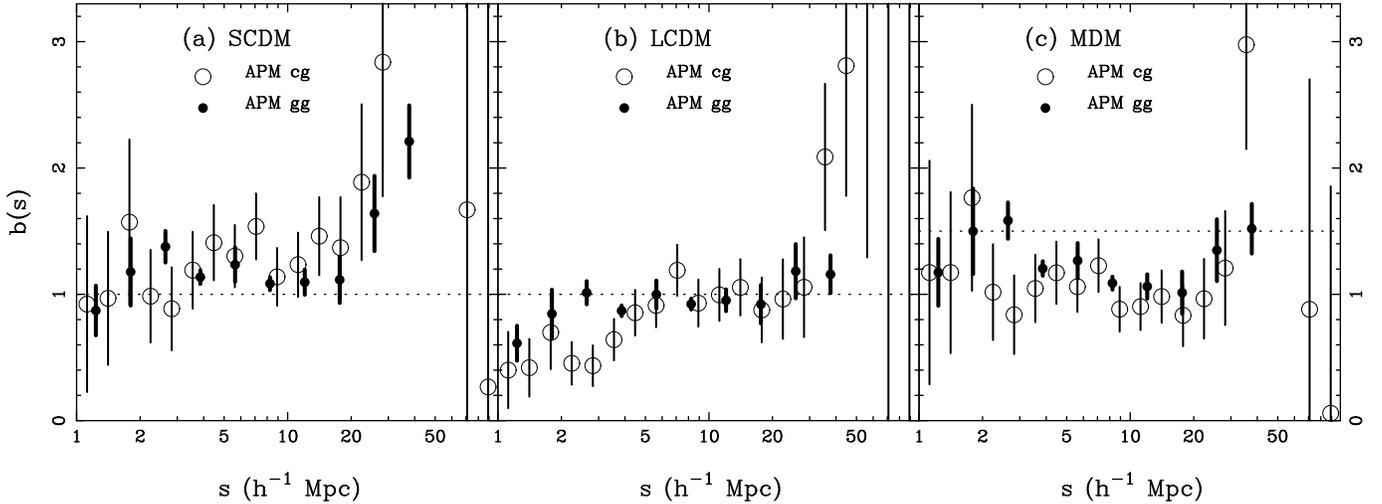}
\caption[bias]{
The bias $b(s)$ as a function of scale in redshift space for 3 cosmological 
models. $b(s)$ is found by dividing the APM $\xi_{cg}(s)$ by  $\xi_{c\rho}(s)$
for each model, the results being represented by open circles in each panel.
A separate estimate of $b(s)$ can be found by evaluating $\sqrt{\xi_{gg}(s)/
\xi_{\rho\rho}}$ and this quantity is plotted as filled circles. 
The dotted line shows in each case the relation that would be expected 
if there exists a linear biasing relation between galaxies and mass and 
$\sigma_{8}$ for galaxies $=1.0$.
 \label{bias}}
\end{figure*}

To calculate how biasing of galaxy fluctuations is expected to change with
scale for each cosmological model, we calculate the particle-particle
two point correlation function, $\xi_{\rho\rho}(r)$ as well
as the cluster-particle cross-correlation function $\xi_{c\rho}(r)$.
Given the observational results for  $\xi_{gg}(r)$ and $\xi_{cg}(r)$,
we can define two biasing relations for each cosmological model, 
\begin{equation}
b^{2}_{gg}(r)=\xi_{gg}(r)/\xi_{\rho\rho}(r)
\label{biasgal}
\end{equation}
and
\begin{equation}
\label{biascg}
b_{cg}(r)=\xi_{cg}(r)/\xi_{c\rho}(r).
\end{equation}
If the efficiency of galaxy formation is not changed by proximity to
clusters, then these quantities, $b_{gg}(r)$ and $b_{cg}(r)$, will be
equal.  We use the estimates of $\xi_{gg}(s)$ calculated from the
APM-Stromlo redshift survey of Loveday \etal (1992) in Equation 14.
As the observed estimates of $\xi{gg}$ and $\xi_{cg}$ are computed in
redshift-space, we compare them with redshift-space estimates of
$\xi_{\rho\rho}$ and $\xi_{c\rho}$ computed from the N-body
simulations.  The results for the three different models are shown in
Figure 14.  In each panel we show both $b_{gg}(s)$ and $b_{cg}(s)$. It
is clear that for SCDM, the required bias must increase strongly
towards larger scales.
For LCDM, the derived   $b_{gg}$ is roughly equal to unity and
nearly independent of scale. 
There does appear to be a discrepancy between $b_{cg}(s)$ and
$b_{gg}(s)$ on small scales, but it is likely that errors in the
cluster redshifts have suppressed the small-scale amplitude of
$\xi_{cg}(s)$ leading to an underestimate of $b_{cg}$.  The MDM model
requires scale-dependent biasing, at least on scales $\simlt 5 \hmpc$.
The dashed lines in the figures show the expectations of linear theory
and constant biasing for models normalised to the COBE measurements
according to the linear theory value of $\sigma_{8}$. In the case of
LCDM, this agrees well with the non-linear simulations, although for
MDM we can see that non-linear evolution appears have made the matter
fluctuation amplitude higher, and less biasing is necessary than
linear theory predicts. 

\begin{figure*}
\centering
\vspace{7.5cm}
\includegraphics{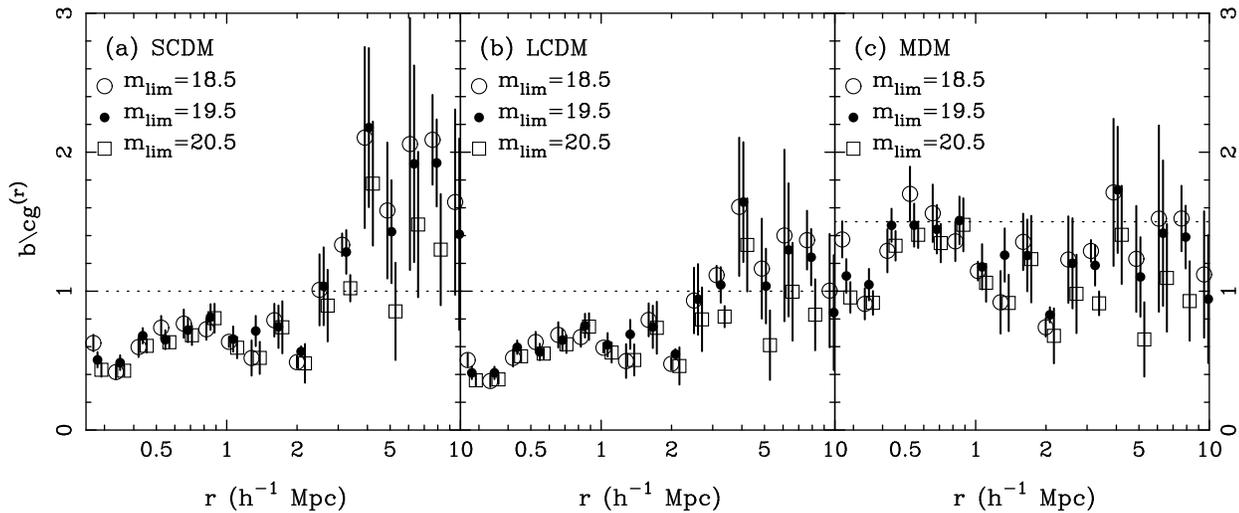}
\caption[biass]{
The bias $b_{cg}(r)$ as a function of scale in real space for 3 cosmological 
models for the region within $10 \hmpc$ of cluster centres.
 b(r) is found by dividing the APM $\xi_{cg}(r)$ )
by  $\xi_{c\rho}(r)$. The real space cross-correlation function 
is found  by inverting the projected cross-correlation 
function of APM galaxies and clusters (see Section 2.2).
 Results for three different galaxy 
magnitude limits are shown, $m_{lim}=18.5$ (filled circles),
$m_{lim}=19.5$ (open circles) and $m_{lim}=20.5$ (open squares).
The amplitude of the models has been adjusted to 
our usual COBE normalisation so that $\sigma_{8}=1.0$ for panels (a) and
(b) and $\sigma_{8}=0.67$ for panel (c).
The dotted line shows in each case the result for $b_{cg}(r)$ 
that would be expected 
if there exists a linear biasing relation between galaxies and mass and if 
$\sigma_{8}$ for galaxies $=1.0$.
 \label{biass1}}
\end{figure*}

 On smaller scales ($s\leq 4 \hmpc$), we have seen that $b_{cg}(s)$
appears to be systematically lower than $b_{gg}(s)$ for LCDM and, to a
lesser extent, for MDM. As mentioned above, errors in cluster
redshifts could be causing this effect. Furthermore, the relation
between $b(s)$ in redshift-space and the underlying $b(r)$ depends on
the velocity field in each model and so mixes physical scales in a
complex way. We will therefore turn to the real space information that
we have available, in the form of $\xi_{cg}(r)$ estimated by inversion
of the projected APM cluster-galaxy cross-correlation function
(Section 2.2). Recovery of spatial correlation functions from
projected and angular statistics can depend sensitively on the
luminosity function assumed. In Figure 15 we plot $b_{cg}(r)$ obtained
when three different magnitude limits are applied to the angular
galaxy catalogue (note the difference in the scale of the absicissa
compared to Figure 14). On the small scales we are considering here,
there is not much difference between the results and we can see that
both the CDM models appear to require some anti-biasing.  This is in
agreement with the study made by Jenkins \etal (1998) who derive an
estimate of  bias in models as a function of scale in real space
from galaxy-galaxy clustering (using Equation~\ref{biasgal}).  Jenkins
\etal use a critical density CDM model which has a lower amplitude,
and so although that model requires $b_{gg}(r)> 1$ on all scales,
there is a relative antibias on small scales compared to large.  Our
MDM models are reasonably consistent with a constant biasing factor,
consistent with the results shown in Figure 5 .  Of course
as no ``hot'' particles have been used to simulate the neutrinos
directly in these models, there is some uncertainty in the accuracy of the 
cluster profiles on small scales.

It should be added that the models which require
anti-biasing do so on scales which are too small to be probed by our
non-linear infall model ($r \simlt 2 \hmpc$). On larger scales, though,
redshift distortions
of $\xi_{cg}$ can give us more information about biasing in our models.
For example, we would expect the bias factors that result from Equations
~\ref{biasgal} and \ref{biascg}
 to be the same when measured in real and redshift space, at least on
large scales where linear theory should hold. On examination
of Figures 14 and 15, this does appear to be the case for LCDM, but there is
the suggestion that the bias factor for both SCDM and MDM is higher
in real space than redshift space. This can also be
seen from the plot of  $\xi_{cg}$ shown directly in Figure 5. 
This is presumably because the coherent peculiar velocities in these two
models which boost $\xi_{cg}$ in redshift space (Kaiser 1987) are too
high. This is just another way of saying that the amplitudes
of our models (which are COBE normalised) are probably wrong. Lower amplitudes
for these two models, such as those given by using a normalisation
which gives the correct space density of galaxy clusters (White,
Efstathiou \& Frenk 1993, Eke \etal 1996) would presumably match the data
better. Changing the normalization would of course not remedy the fact that
SCDM does not match the shape of clustering on large scales.

\begin{figure*}
\centering
\vspace{7.5cm}
\includegraphics{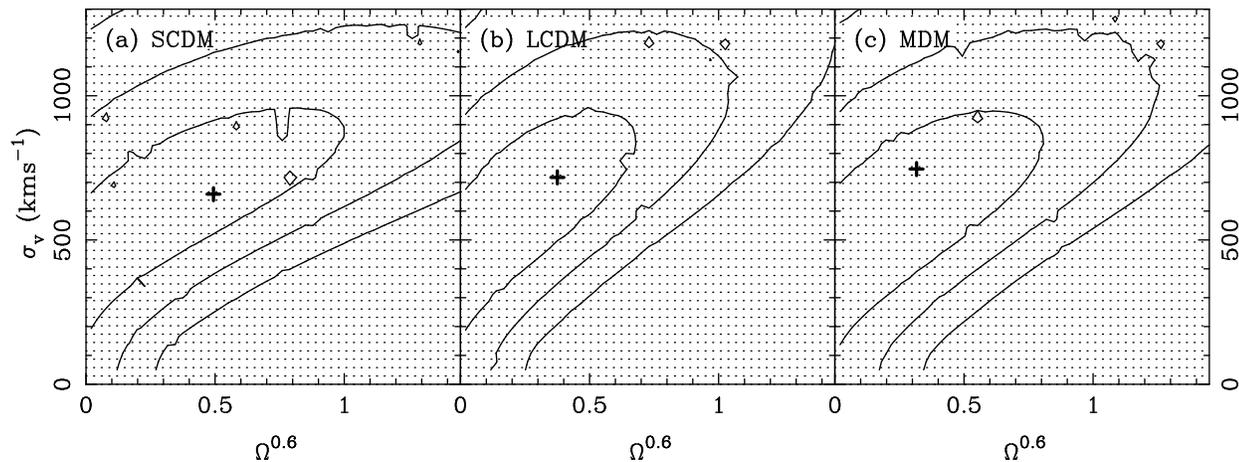}
\caption[modapm]{$\Delta \chi^{2}$ distributions
for $\Omega^{0.6}$ and $\sigma_{v}$ evaluated in a similar way to those
in Figure 13, but this time using the velocity field predicted from the 
mass distribution in models  to distort the real
space $\xi_{cg}$. This will give an estimate of the values
of  $\Omega^{0.6}$ and  $\sigma_{v}$ needed in the models to account for
the redshift distortions, given that they are automatically
correctly biased to give the same real space $\xi_{cg}$ as the 
APM survey.
 \label{biass}}
\end{figure*}

A potentially more powerful way of comparing the models and
observations involves using the full $\xi_{cg}(\sigma,\pi)$
information.  To do this, we first make the assumption that some form
of scale-dependent biasing holds so that $\xi_{cg}(r)$ in the model
being tested and the observations are related by
Equation~\ref{biascg}.  We then assume that the velocity field around
clusters can be predicted from $\xi_{c\rho}(r)$ in the model, and that
the observed form of $\xi_{cg}(\sigma,\pi)$  has been distorted by this
velocity field. We use Equation~\ref{convolve} to model the
distortion, with the non-linear infall prediction being computed from
$\xi_{c\rho}(r)$, and used to distort $\xi_{cg}(r)$.  At this point,
we are guaranteed by construction that the amplitude and shape of the
scale-dependent biasing is correct, but using different values of
$\Omega$ and $\sigma_{v}$ in Equation~\ref{convolve} will result in
differing degrees of agreement with the observed
$\xi_{cg}(\sigma,\pi)$.  We carry out $\chi^{2}$ fitting of
$\xi_{cg}(\sigma,\pi)$ exactly as in Section 4.3, except that we are
constraining $\Omega^{0.6}$ required in each specific model, rather
than $\beta$ for a general scenario with scale-independent biasing.

The results of this procedure are shown in  Figure 16, where we can see that
all models prefer a relatively low value of $\Omega^{0.6}$, which of course
is natural for LCDM but not for the others.
After marginalising over all values of $\sigma_{v}$, we find
the following values for $\Omega^{0.6}$ (with $95\%$ upper limits),
0.49 (1.28), 0.38 (0.88) and 0.32 (1.06), for SCDM, LCDM and
MDM respectively.  Given the tendency of the spherical infall model to slightly
overestimate  $\Omega^{0.6}$, these results are reasonably significant.
The $\sigma_{v}$ results for all models are $\sim 700 \kms$. As before,
we can attribute this high value to the large errors in the redshifts of
individual APM clusters which should bias the $\sigma_{v}$ results upwards.
It is interesting that the SCDM model and to a lesser extent the MDM model
are expected to give  $\sigma_{v}$ almost this high without accounting for the
uncertainties in the velocities. As we know that these errors are present
(Dalton \etal 1994a), we can regard 
this is additional evidence against these two models.

In this analysis, we have two parameters, $\sigma_{v}$ and
$\Omega^{0.6}$, each of which probes a different part of the velocity
field around clusters. If the model is correct then the best fit
values from the observations should match the values of both
parameters predicted from the model. We have seen that this fitting procedure
favours the LCDM model over the other two we consider here. However,
we have been restricted to a narrow set of model parameters
and it is possible that by varying these it might be possible to
improve the agreement with observations. For example, a
lower amplitude SCDM model would probably require a higher value of
$\Omega$ to fit the data and so be more consistent.  

This sort of analysis could also be applied to the galaxy-galaxy
correlation function, although the results would probably not be as
reliable. The analysis of Section 4 shows that non-linear dynamical
model of cluster-galaxy peculiar velocities works accurately even on
small scales. However, it has proved difficult to develop a model of
galaxy-galaxy velocities of comparable accuracy (see e.g. Hatton and
Cole 1997). Throughout our analysis we have neglected the possibility
of  velocity biasing. Velocity biasing could, if significant,
introduce systematic errors in analyses of  redshift-space
distortions. Unfortunately, theoretical predictions for
velocity bias are so uncertain  ({\it e.g.} Summers, Davis and Evrard 1995)
that it is not yet possible to make reliable models of this effect.

\section{Discussion and Conclusions}

The analysis presented in this paper  differs  from other recent work
on the density and velocity fields around clusters in a number of
ways.  The CNOC cluster redshift survey of Carlberg \etal (1997)
specifically targets galaxies within and close to the virialised region
of clusters. Carlberg \etal find that a rescaling of cluster profiles 
allows all clusters to be fit by the universal dimensionless halo
profile of Navarro, Frenk and White (1997). We do not carry out such
a rescaling here, as we are interested in the profiles of clusters
out to a much larger distance (where presumably rescaling would not work,
as all profiles must converge to the mean density). Our dynamical measure
of the dark matter in each cluster  is also based on a different velocity
and density regime, the infall region, rather than the virialised region.

The sort of analysis we employ here could be extended to constrain
$\Omega$ from the virialised region. 
In the usual analysis of cluster velocity dispersions, there is no
way of telling unambiguously whether a given galaxy is a cluster member
with a high velocity or a foreground or background galaxy. Non-member
galaxies and the various ways employed to prune them
can add significant bias  to determinations of the velocity dispersions 
from samples of clusters (see e.g. van Haarlem \etal 1997).
Analysis of the ``finger of God'' effect on the cross-correlation
function offers a way of treating the problem statistically, which 
automatically accounts for contamination. In this paper,  we have not
made  use of  the information on $\sigma_{v}$ because we  have a
relatively  small number of galaxies in our sample
and because there are uncertainties in the errors of the APM cluster
redshifts.
Future surveys  such as the 2dF and Sloan redshift surveys
(Colless 1997, Gunn \& Weinberg 1995)
 and also the presently available Las Campanas Redshift survey 
(Schectman \etal 1996) could be used effectively 
for this purpose.  These large surveys
will also form good datasets to which we can apply the non-linear infall 
models used in this paper, and should  provide accurate estimates of
$\Omega$ and biasing parameters from the infall region of clusters. 

The value of $\beta$ we obtain in this paper is consistent with that
obtained from a consideration of linear redshift distortions of the
galaxy-galaxy clustering function on larger scales (e.g. Loveday \etal
1996). This suggests that any significant biasing as a function of
scale is probably confined to scales smaller than $\sim 2 \hmpc$.

In summary, we have investigated the shape and amplitude of cluster
profiles both in the APM survey and in simulations of cosmological
models with Gaussian initial conditions. In all cases we find a
distinct two-component structure, with the shape of 
$\xi_{cg}(r)$ on large scales consistent with the
galaxy-galaxy or particle-particle correlation functions. For $r\simlt
5 \hmpc$ we see a steepening of the profile. This region is also the
part most sensitive to differences in the richnesses of clusters used
to calculate $\xi_{cg}(r)$, although the effects are too small to be
noticeable in the observational sample.

As expected for objects forming in a bottom-up scenario, the velocity
and density field around clusters in the simulations is complex, with
substructure evident in plots. Nevertheless, when we average the
infall velocities around all clusters, the shape and amplitude of the
infall curve is surprisingly well described by a spherically symmetric
collapse model.  It is therefore feasible to use this simple velocity
model to describe distortions in $\xi_{cg} (\sigma,\pi)$ and therefore to
constrain $\Omega$.  When we do this using the APM data, we find that
$\Omega^{0.6}/b$ has a best fit value of 0.43 and is constrained to be
less that 0.87 at $95 \%$ confidence, where $b$ is the linear bias
parameter.

We have also compared $\xi_{gg}$ and $\xi_{cg}$ with N-body simulations of
the mass distribution to estimate the biasing of galaxy fluctuations
as a function of scale. We conclude that anti-biasing on
scales $r \simlt 2\hmpc$ is required in the SCDM and LCDM senarios, as
their cluster profiles appear to be steeper than in the observations.
This sort of biasing might be possible, as galaxies or quasars forming
in clusters might suppress galaxy formation nearby (e.g. Babul \&
White 1991). On the other hand, this might be hard to
reconcile with a galaxy density field which is totally unbiased on
large scales (as may be required in  LCDM).  The MDM model does
have the correct cluster profiles, without any scale-dependent
biasing. However, we find that the redshift distortions in the
cross-corrrelation function predicted by this model are too strong, by
about $2 \sigma$, so that the LCDM model should still be considered
the most successful of those tested.

Application of the techniques presented in this paper to larger redshift
 surveys should yield robust measures of  the cosmic density from the
infall region around galaxy clusters. The spherical infall model is 
a dynamical approximation which is accurate in the highly non-linear regime
and should continue being useful for constraining  $\Omega$ and
biasing in a fashion complementary to traditional linear analyses of 
galaxy-galaxy clustering.

\section*{Acknowledgments}
 This work was supported by grants
from the UK Particle Physics and Astronomy Research Council, and by NASA 
Astrophysical Theory Grants NAG5-2864 and NAG5-3111. GPE acknowledges
the award of a  PPARC Senior Fellowship.

\bigskip

\setlength{\parindent}{0mm}
{\bf REFERENCES} 
\bigskip

\def\refe {\par \hangindent=.7cm \hangafter=1 \noindent}
\def\apj { ApJ }
\def\mn { MNRAS }
\def\apl { ApJ Lett.}
\def\aa { A\&A}

\refe Babul, A. \& White, S. D.~M., 1991, \mn, {\bf 253}, 31P.

\refe Bardeen J.M., Bond J.R., Kaiser N. \& Szalay A., 1986, \apj, 
{\bf 304}, 15.

\refe Carlberg, R.~G., Yee, H. K.~C., Ellingson, E., Morris, S.~L.,
Abraham, R., Gravel, P., Pritchet, C.~J., Smecker-Hane, T., Hartwick,
F.D.~A., Hesser, J.~E., Hutchings, J.~B., Oke, J.~B., 1997,
 \apl, {\bf 485}, L13.
 
\refe Colberg, J.~M., White, S. D.~M., Jenkins, A., \& Pearce, F.~R.,
1997, \mn, {\it submitted}, astro-ph/9711040.

\refe Colless, M., 1997, in {\it Wide Field  Spectroscopy},
 eds Kontizas M., Kontizas E., Kluwer. 

\refe Croft, R. A.~C. \&  Efstathiou, G., 1994, \mn, {\bf 267}, 390. 
 
\refe Dalton, G.~B., 1992, {\it D.Phil thesis}, Oxford University.
 
\refe Dalton, G.~B., Efstathiou, G., Maddox, S.~J.  \& Sutherland, W.~J., 1992.
\apl, {\bf 390}, L1.

\refe Dalton, G.~B., Efstathiou, G., Maddox, S.~J.  \& Sutherland, W.~J.,
1994a. \mn, {\bf 269}, 151.

\refe Dalton, G.~B., Croft, R. A.~C.,  Efstathiou, G., Sutherland, W.~J., 
 Maddox, S.~J.  \& Davis, M., 1994b.
\mn, {\bf 271}, L47.

\refe Dalton, G.~B., Efstathiou, G., Maddox, S.~J.  \& Sutherland, W.~J.,
1997, \mn, {\bf 289}, 263.

\refe Davis, M. \& Peebles, P. J.~E., 1977, \apj, {\bf 34}, 425.

\refe Diaferio, A. \& Geller, M.~J., 1997, \apj, {\bf 481}, 633.

\refe Efstathiou, G., Davis, M., Frenk, C.~S., \& White, S.D.~M., 1985,
{\it Ap. J. Suppl.}, {\bf 57}, 241.
 
\refe Efstathiou, G., 1988, In: {\it Comets to Cosmology}, 312, ed. Lawrence,
 A., Springer  Verlag.
 
 
\refe Efstathiou, G., 1993, In: {\it Proceedings of the U.S. National Academy
 of Sciences}, 4859,
  ed. 90.
 
\refe Eke, V.~R., Cole, S. \& Frenk, C.~S., 1996., \mn, {\bf 282}, 263.

\refe Gazta{\~n}aga, E., 1995, \apj, {\bf 454 }, 561

\refe Gunn, J., Weinberg, D. H., 1995, in {\it Wide-Field Spectroscopy and
the Distant Universe}, eds S.J. Maddox and A. Arag\'{o}n-Salamanca,
World Scientific, Singapore.
 
\refe Hatton, S.~J. \& Cole, S., 1997, \mn, {\it submitted}, astro-ph/9707186.

\refe Hoffman, Y., 1986.
 \apj, {\bf 308}, 493.
 
\refe Jenkins, A., Frenk, C.~S., Thomas, P.~A., Colberg, J.~M., White,
S.D.~M., Couchman, H.M.~P., Peacock, J.~A., Efstathiou, G., Nelson, A.~H.,
1998, \apj, {\it in press}, astro-ph/9709010.

\refe Kaiser, N., 1987.
 {\it Mon. Not. R. astr. Soc.}, {\bf 227}, 1.
 
\refe Klypin, A., Holtzman, J., Primack, J.  \& Reg\"{o}s, E., 1993,
 \apj, {\bf 416}, 1.

\refe Lahav, O., Rees, M.~J., Lilje, P.~B., Primack, J.~R., 1991,
 {\it Mon. Not. R. astr. Soc.}, {\bf 251}, 128.
 
\refe Lilje, P.~B. \& Efstathiou, G., 1988,
 \mn, {\bf 231}, 635.
 
\refe Lilje, P.~B. \& Efstathiou, G., 1989, \mn, {\bf 236}, 851.
 
\refe Lilje, P.~B. \& Lahav, O., 1991, \apj, {\bf 374}, 29.

\refe Loveday, J., 1990, {\it PhD thesis}, Cambridge University.
 
\refe Loveday, J., Peterson, B.~A., Efstathiou, G.,  \& Maddox, S.~J., 1992a,
 {\it Ap. J. Lett.}, {\bf 390}, 338.

\refe Loveday, J., Efstathiou, G., Peterson, B.~A.  \& Maddox, S.~J., 1992b,
 {\it Ap. J. Lett.}, {\bf 400}, L43.

\refe Loveday, J., Efstathiou, G.,  Maddox, S.~J. \& Peterson, B.~A., 1996,
 \apj, {\bf 468}, 1.

\refe Maddox, S.~J., Efstathiou, G., Sutherland, W.~J., \& Loveday, J., 1990a,
\mn, {\bf 242}, 43.

\refe Maddox, S.~J., Efstathiou, G., Sutherland, W.~J., \& Loveday, J., 1990b,
\mn, {\bf 243}, 692.

\refe Maddox, S.~J., Efstathiou, G., Sutherland, W.~J. 1996,
\mn, {\bf 283}, 1227.

\refe Merch\'{a}n, M.~E., Abadi, M.~G., Lambas, D.~G., \& Valotto, C., 1997,
\apj, {\it in press}.

\refe Mo, H.~J., Peacock, J.~A., \& Xia, X.~Y., 1993, \mn, {\bf 260},121.
 
\refe Moore, B., Frenk, C.~S., Efstathiou, G.  \& Saunders, W., 1994,
\mn, {\bf 269}, 742.
 
\refe Navarro, J.~F., Frenk, C.~S., \& White, S. D.~M., 1997, \apj, {\it
in press}, astro-ph/9611107.

\refe Peebles, P. J.~E., 1980, 
{\it The Large-Scale Structure of the Universe}, Princeton University Press.

\refe Peebles, P. J.~E., 1993, 
{\it Principles of Physical Cosmology}, Princeton University Press.
 
\refe Ratcliffe, A., Shanks, T., Parker, Q.~A., Fong, R., 1997,
 \mn, {\it in press}, astro-ph/9702227.

\refe Reg\``{o}s, E. \& Geller, M.~J., 1989,
 {\it Astron. J.}, {\bf 98}, 755.

\refe Saunders, W., Rowan-Robinson, M. \& Lawrence, A., 1992, \mn, 258, 134.
 
\refe Schectman, S.~A., Landy,S.~D.,Oemler, A., Tucker, D.~L., Lin, H.,
Kirshner, R.~P., Schecter, P.~L., 1996, {\bf 470}, 172. 

\refe Seldner, M. \& Peebles, P. J.~E., 1977,
 \apj, {\bf 215}, 703.
 
\refe Summer, F.~J., Davis, M. \& Evrard, A., 1995, \apj, {\bf 454},1.

\refe van Haarlem, M., 1992,{\it PhD thesis}, Leiden University.

\refe van Haarlem, M.,Frenk, C.~S. \& White, S.D.~M., 1997, \mn,
{\bf 287}, 817.

\refe van Haarlem, M., Cayon, L., de~la Cruz, C.~G., Martinez-Gonzalez, E.  \&
  Rebolo, R., 1993, \mn, {\bf 264}, 71.
 
\refe van Haarlem, M. \& van~de Weygaert, R., 1993, \apj, {\bf 308}, 499.
 
\refe Villumsen, J.~V. \& Davis, M., 1986,
 \apj, {\bf 308}, 499.

\refe Weinberg, D.~H.,  in {\it Wide-Field Spectroscopy and the
Distant Universe, eds. S. J. Maddox and A. Arag\'on-Salamanca,
(Singapore: World Scientific)} , 129, astro-ph/9409094 
 
\refe White, S.D.~M., Efstathiou, G. \& Frenk, C.~S., 1993,
\mn, {\bf 262}, 1023.

\refe Wright, E.~L., Smoot, G.~F., Kogut, A., Hinshaw, G., Tenorio,
 L., Lineweaver,  C., Bennett, C.~L.  \& Lubin, P.~M., 1994,
 \apj, {\bf 420}, 1.
 
\refe Yahil, A., Sandage, A.  \& Tammann, G.~A., 1980,
 \apj, {\bf 242}, 448.
 
\refe Yahil, A., 1985.
 In: {\it The Virgo Cluster of Galaxies}, 359, eds Richter, O.~G. \&
  Binggeli, B., ESO : Garching.

\end{document}